\documentclass[preprints,article,accept,pdftex,moreauthors]{Definitions/mdpi} 
\firstpage{1} 
\pubvolume{1}
\issuenum{1}
\articlenumber{0}
\pubyear{2022}
\copyrightyear{2022}
\datereceived{} 
\dateaccepted{} 
\datepublished{} 
\hreflink{https://doi.org/} % If needed use \linebreak
\usepackage{qting}
\renewcommand{\hl}[1]{#1}

\usepackage{GB_glossaries}

\newacronym[type=symbolslist]{TCom}{\ensuremath{\widetilde{T}}}{Green's function for the complex Transmittance}

\newacronym[type=symbolslist]{SRTCom}{\ensuremath{\text{SR}\{\acrshort{TCom}\}}}{Single-Ratio of the \gls{TCom}}
\newacronym[type=symbolslist]{SRTComI}{\ensuremath{|\text{SR}\{\acrshort{TCom}\}|}}{\gls{SRTCom} amplitude}
\newacronym[type=symbolslist]{SRTComLnI}{\ensuremath{\ln|\text{SR}\{\acrshort{TCom}\}|}}{natural logarithm of \gls{SRTComI}}
\newacronym[type=symbolslist]{SRTComP}{\ensuremath{\angle\text{SR}\{\acrshort{TCom}\}}}{\gls{SRTCom} phase}

\newacronym[type=symbolslist]{DRTCom}{\ensuremath{\text{DR}\{\acrshort{TCom}\}}}{Dual-Ratio of the \gls{TCom}}
\newacronym[type=symbolslist]{DRTComI}{\ensuremath{|\text{DR}\{\acrshort{TCom}\}|}}{\gls{DRTCom} amplitude}
\newacronym[type=symbolslist]{DRTComLnI}{\ensuremath{\ln|\text{DR}\{\acrshort{TCom}\}|}}{natural logarithm of \gls{DRTComI}}
\newacronym[type=symbolslist]{DRTComP}{\ensuremath{\angle\text{DR}\{\acrshort{TCom}\}}}{\gls{DRTCom} phase}

\newacronym[type=symbolslist]{CCom}{\ensuremath{\widetilde{C}}}{complex optical Coupling, power, and/or efficiency factor}
\glsdisablehyper

\newacronym[type=productlist]{SS150H}{SphereSpectro}{Gigahertz Optik SphereSpectro 150H (T\"uerkenfeld, Germany)}
\newacronym[type=productlist]{PE1050}{LAMBDA 1050}{Perkin Elmer LAMBDA 1050+ (Waltham, MA USA)}
\newacronym[type=productlist]{PE365}{LAMBDA 365}{Perkin Elmer LAMBDA 365+ (Waltham, MA USA)}

\usepackage{siunitx}
\usepackage{threeparttable}
\usepackage{matlab-prettifier}
\usepackage{dirtytalk}

\makeatletter
\def\UrlAlphabet{%
      \do\a\do\b\do\c\do\d\do\e\do\f\do\g\do\h\do\i\do\j%
      \do\k\do\l\do\m\do\n\do\o\do\p\do\q\do\r\do\s\do\t%
      \do\u\do\v\do\w\do\x\do\y\do\z\do\A\do\B\do\C\do\D%
      \do\E\do\F\do\G\do\H\do\I\do\J\do\K\do\L\do\M\do\N%
      \do\O\do\P\do\Q\do\R\do\S\do\T\do\U\do\V\do\W\do\X%
      \do\Y\do\Z}
\def\UrlDigits{\do\1\do\2\do\3\do\4\do\5\do\6\do\7\do\8\do\9\do\0}
\g@addto@macro{\UrlBreaks}{\UrlOrds}
\g@addto@macro{\UrlBreaks}{\UrlAlphabet}
\g@addto@macro{\UrlBreaks}{\UrlDigits}
\makeatother

\Title{Method for %MDPI: In order to process this paper smoothly, please check if anything is missing in the main text and offer responses to these comments, and please do not delete them. GB: OK
 Measuring Absolute Optical Properties of Turbid Samples in a Standard Cuvette}
\TitleCitation{Method for Measuring Absolute Optical Properties of Turbid Samples in a Standard Cuvette}

\Author{\hl{Giles Blaney} %MDPI: Please carefully check the accuracy of names and affiliations. GB: Done
 *\orcidA{}, Angelo Sassaroli \orcidB{} and Sergio Fantini \orcidC{}}
\AuthorNames{Giles Blaney, Angelo Sassaroli, and Sergio Fantini}

\AuthorCitation{Blaney, G.; Sassaroli, A.; Fantini, S.}
\address[1]{\hl{Department of Biomedical Engineering, Tufts University}, %MDPI: Addresses should be sorted from subordinate to superior, so we revised the order, please confirm. GB: Okay, also added GB's adress
 4 Colby Street, \mbox{Medford, MA 02155, USA}; giles.blaney@tufts.edu (G.B.); angelo.sassaroli@tufts.edu (A.S.); sergio.fantini@tufts.edu (S.F.)}

\corres{\hangafter=1 \hangindent=1.05em \hspace{-0.82em} Correspondence: giles.blaney@tufts.edu}

\abstract{{Many applications seek to measure a sample's absorption coefficient spectrum to retrieve the chemical makeup.} Many real-world samples are optically turbid, causing scattering confounds which many commercial spectrometers cannot address.
\q{R1C1}{{Using diffusion theory and considering absorption and reduced scattering coefficients on the order of \SI{0.01}{\per\milli\meter} and \SI{1}{\per\milli\meter}, respectively, we develop a method which utilizes frequency-domain to measure absolute optical properties of turbid samples in a standard cuvette (\SI{45x10x10}{\milli\meter}).}}
Inspired by the self-calibrating method, which removes instrumental confounds, the method uses measurements of the diffuse complex transmittance at two sets of two different source-detector distances. {We find: this works best for highly scattering samples (reduced scattering coefficient above \SI{1}{\per\milli\meter}); higher relative error in the absorption coefficient compared to the reduced scattering coefficient; accuracy is tied to knowledge of the sample’s index of refraction.} Noise simulations with \SI{0.1}{\percent} amplitude and $\ang{0.1}=\SI{1.7}{\milli\radian}$ phase uncertainty find errors in absorption and reduced scattering coefficients of \SI{4}{\percent} and \SI{1}{\percent}, respectively. We expect that higher error in the absorption coefficient can be alleviated with highly scattering samples and that boundary condition confounds may be suppressed by designing a cuvette with high index of refraction. {Further work will investigate implementation and reproducibility.}}%200/200 words maximum

\keyword{{{absolute optical properties; absorption coefficient; reduced scattering coefficient; diffusion}} {{theory; turbid samples; optical spectroscopy; sample measurement; cuvette; frequency-domain near-}} {infrared spectroscopy; self-calibration}}%List three to ten pertinent keywords specific to the article; yet reasonably common within the subject discipline. 

\featuredapplication{Quantitative and calibration free determination of the absolute optical properties of turbid samples in a standard cuvette with milliliter scale volume.}
\begin{document}
%%%%%%%%%%%%%%%%%%%%%%%%%%%%%%%%%%%%%%%%%%
\section{Introduction}
Samples in which the propagation of light is dominated by random scattering are considered optically diffuse. Such samples can be characterized by two absolute optical properties, the~{\gls{mua}}  and the \gls{musp} \cite{bigioQuantitativeBiomedicalOptics2016}. The~\gls{mua} represents chemical information, and~its spectral measurement allows for determination of the sample's chemical constituents and concentrations. Meanwhile, the~\gls{musp} describes the \si{micrometer} scale structure of diffuse samples. However, in~many applications \gls{musp} is considered a confounding parameter, since measurement of \gls{mua} and chemical makeup is often the end goal. For~this reason, even when diffuse sample measurement of only \gls{mua} is sought, \gls{musp} must also be determined since it significantly impacts the behavior of light and thus the recovered \gls{mua}. \par

Applications that seek to measure these diffuse optical properties are numerous and span many fields. For example, applications include those within biomedical research and clinical applications~\cite{bigioQuantitativeBiomedicalOptics2016,quaresimaFunctionalNearInfraredSpectroscopy2019,rahmanNarrativeReviewClinical2020}, of~food science and quality~\cite{ozakiNearInfraredSpectroscopyFood2006, johnsonOverviewNearinfraredSpectroscopy2020, kademiApplicationsMiniaturizedPortable2019}, concerning pharmaceutical metrology~\cite{razucNearinfraredSpectroscopicApplications2019, stranzingerReviewSensingTechnologies2021}, pertaining to art and archaeology~\cite{chenAuthenticationInferenceSeal2017, trantVisibleNearinfraredSpectroscopy2020}, and~within dendrology~\cite{tsuchikawaReviewRecentApplication2015} to name a few. In~all cases, one has two~options:
\begin{enumerate}
	\item To make a measurement which retrieves the \gls{mut} or the \gls{mueff}. \label{enum:sep_muaORmusp}
	\item To make a measurement that can separate both \gls{mua} and \gls{musp}. \label{enum:sep_muaANDmusp}
\end{enumerate}

However, only option~\ref{enum:sep_muaANDmusp} allows for careful quantitative analysis of the sample properties; since in option~\ref{enum:sep_muaORmusp} one can only measure a coefficient, namely \gls{mut} or \gls{mueff}, that couples both the \gls{mua} and \gls{musp} of the sample. There are few methods capable of achieving option~\ref{enum:sep_muaANDmusp}. 
\q{R2C1a}{{One such technique is \gls{NIRS} implemented in \gls{FD} \cite{fantiniFrequencyDomainTechniquesCerebral2020} (or \gls{TD} \cite{torricelliTimeDomainFunctional2014}) which can recover \gls{mua} and \gls{musp} by using temporally modulated light. In~the case of \gls{FD}, photon density waves are generated by using a sinusoidally modulated source on the order of \SI{100}{\mega\hertz}, and~the amplitude and phase of these photon density waves are measured from the detected modulated light signal. An~example of a commercially available \gls{FD} instrument capable of this measurement is the \gls{ISSv2}.}}
A second technique capable of option~\ref{enum:sep_muaANDmusp} is the integrating sphere~\cite{foschumPreciseDeterminationOptical2020, bergmannPreciseDeterminationOptical2020}. This technique measures total diffuse reflectance and total diffuse transmittance to separate \gls{mua} and \gls{musp}. Both techniques have their strengths and weaknesses.
\q{R1C2a}{{Common implementations of \gls{FD} \gls{NIRS} require large sample volumes to create geometries that are effectively infinite in at least one dimensional extent making simple diffusion theory expressions valid~\cite{continiPhotonMigrationTurbid1997}. Other methods, besides~diffusion theory, exist to tackle non-simple geometries such as Monte Carlo~\cite{fangMonteCarloSimulation2009} and the radiative transport equation implemented with higher order spherical harmonics~\cite{hielscherSagittalLaserOptical2004, kloseInverseSourceProblem2005, kloseLightTransportBiological2006}. However these methods are computationally costly compared to diffusion theory and would likely be impractical to implement on a broad range of \glspl{lam}. Another weakness of \gls{FD} regarding the number of \glspl{lam}, is the typical implementation at only discrete \glspl{lam} such as with the \gls{ISSv2}.}}
Meanwhile, the~integrating sphere requires careful calibration or a reference sample and is easily susceptible to errors induced by the measurement technique (for example, light loss causing an incorrect measurement of total reflectance and transmittance). \par

Due to these difficulties with option~\ref{enum:sep_muaANDmusp} (and relative ease implementing option~\ref{enum:sep_muaORmusp}), most commercial spectrometers make a measurement that is based on the retrieving non-diffuse transmittance. Therefore, for~quantitative determination of a sample's chemical concentrations (through the \gls{mua} spectrum), samples must be non-scattering or transparent; either innately or through some chemical washing. Whenever this is not possible the measurement will be confounded by scattered light. This implies that the \gls{mua} will be overestimated and its spectral dependence distorted leading to errors in the estimation of a sample's chemical~constituents. \par

{One such instrument that shines when samples are transparent and non-scattering is the \acrlong{PE365}, this and instruments like it are the workhorses of many chemical and biological laboratories.} However, when diffuse sample measurement is necessary (and quantitative measurement of properties sought), one of the aforementioned techniques capable of option~\ref{enum:sep_muaANDmusp} is required. One example is the \gls{SS150H}, an~instrument directly designed for spectroscopic measurement of both \gls{mua} and \gls{musp} via integrating sphere. Furthermore, integrating spheres may be purchased as attachments to traditional spectrometers, thus adding diffuse functionality. One such example of a spectrometer that has this option is the \acrlong{PE1050}. However, we are not aware of any commercially available instrument that utilizes the \gls{FD} in such~applications. \par

{Because of the apparent gap in the market for instruments which complete diffuse measurement of \gls{mua} and \gls{musp}, namely implementation with instruments that utilize \gls{FD} \gls{NIRS} like techniques, we will focus closer on \gls{FD}.}
\q{R2C1b}{{Measurements of \gls{mua} and \gls{musp} with temporally modulated light, such as \gls{FD} \gls{NIRS}, is actually rather common but typically only in the research setting (using the \gls{ISSv2} for example).}}
However, we know of no \gls{FD} instruments designed for the sample sizes and form factors of traditional spectrometers which accept a cuvette.
\q{R1C2b}{{In-fact \gls{FD} \gls{NIRS} methods typically require large sample volumes on the scale of \si{liters} to implement simple diffusion theory solutions.}}
\q{R4C1}{{Two examples of work that considered \gls{FD} measurements in confined regions were that in the slab~\cite{hoMultiHarmonicHomodyneApproach2012} or block~\cite{taniguchiLightDiffusionModel2007} (which utilized the same diffusion theory model implemented here~\cite{kienleLightDiffusionTurbidParallelepiped2005}), however the geometries considered in these works were rather large compared to a cuvette.}}
There are advantages of \gls{FD} \gls{NIRS} which would lead one to seek or design and manufacture such an instrument. For~example, \gls{FD} \gls{NIRS} can leverage existing techniques which eliminate the need for instrumental calibration such as the \gls{SC} method~\cite{hueberNewOpticalProbe1999}. Additionally, despite \gls{FD} typically being implemented at discrete \gls{lam}, methods exist to achieve broadband \gls{mua} measurement. {This may be done by combining measurements at discrete \glspl{lam} in \gls{FD} (or \gls{TD}) with measurements at broadband \glspl{lam} with \gls{CW} \cite{bevilacquaBroadbandAbsorptionSpectroscopy2000,osullivanDiffuseOpticalImaging2012,vasudevanMethodQuantitativeBroadband2020}, implemented with \gls{SC} and a method dubbed \gls{DS}, respectively, \cite{blaneyBroadbandAbsorptionSpectroscopy2021, sassaroliDualslopeMethodEnhanced2019, blaneyPhaseDualslopesFrequencydomain2020}.} This leverages the fact that \gls{SC} and \gls{DS} both rely on a difference type measurement which is capable of subtracting away instrumental~confounds. \par

We see an opportunity to develop a method which leverages the tools available in \gls{FD} \gls{NIRS} to measure absolute \gls{mua} and \gls{musp} in a standard cuvette with \si{milliliter} scale volume in an attempt to compete with the existing integrating sphere type devices. Therefore, in~this work we present a method that utilizes \gls{FD} \gls{NIRS} in a small geometry the size of a standard cuvette (\SI{45x10x10}{\milli\meter}). Our proposal relies on the \gls{SC}/\gls{DS} method to remove a majority of the instrumental confounds.
\q{R1C2c}{{To our knowledge no work has yet leveraged \gls{SC}/\gls{DS} directly on the cuvette geometry as we propose here. First, we utilize a seldom implemented but still computationally inexpensive diffusion theory derived expression for the box geometry~\cite{kienleLightDiffusionTurbidParallelepiped2005} to model our proposed measurement and determine the method's feasibility in theory.}}
Then we further develop ways to retrieve \gls{mua} and \gls{musp} from the proposed measurement. Lastly, we determine the strengths and weaknesses of the proposed method.
\q{R1C2d}{{Our end goal is to implement the method for broadband \gls{lam} measurement of \gls{mua} \cite{blaneyBroadbandAbsorptionSpectroscopy2021}, thus computation cost and model simplicity are of importance leading to the choice of a diffusion theory model.}}
In this article we focus only on the \gls{FD} part since the extension to broadband \gls{lam} \gls{CW} will utilize all the same~theory. \par

%%%%%%%%%%%%%%%%%%%%%%%%%%%%%%%%%%%%%%%%%%
\section{Methods}
\subsection{Geometry}
In this work, we consider a box geometry with the dimensions of a standard cuvette (\SI{45x10x10}{\milli\meter}; Figure \ref{fig:geo}). A~\gls{DS}/\gls{SC} arrangement \hl{(}the word slope in \acrfull{DS} is historical~\cite{sassaroliDualslopeMethodEnhanced2019,blaneyPhaseDualslopesFrequencydomain2020} as no slopes are actually considered in this work\hl{)} %MDPI: footnote is not allowed for this journal, we moved the content to the maintext, please confirm, same below.
% GB: Okay
is achieved by placing \num{2} sources (\texttt{1} \& \texttt{2}; Figure~\ref{fig:geo}{a,b}) and \num{2} detectors (\texttt{A} \& \texttt{B}; Figure~\ref{fig:geo}{b,c}) symmetrically on opposing sides of the cuvette. Using the coordinate system shown in Figure~\ref{fig:geo}, the~optodes were considered at the following \glspl{r}: $\acrshort{r}_{\texttt{1}}=-17 \hat{x}~\si{\milli\meter}$, $\acrshort{r}_{\texttt{2}}=17 \hat{x}~\si{\milli\meter}$, $\acrshort{r}_{\texttt{A}}=-6 \hat{x} +10 \hat{z}~\si{\milli\meter}$, and~$\acrshort{r}_{\texttt{B}}=6 \hat{x} +10 \hat{z}~\si{\milli\meter}$. This forms two possible \glspl{rho} of \SIlist{14.9;25.1}{\milli\meter} (\num{2} each), for~\texttt{1A} \& \texttt{2B} and \texttt{2A} \& \texttt{1B}, respectively. \par

\subsection{Types of~Measurement}
\q{R2C1c}{{The signal obtained between a single temporally modulated source and a single detector recovers the \gls{TCom} with \gls{FD} \gls{NIRS}. \gls{TCom} is a complex number to represent the amplitude and phase of the transmitted photon density waves modulated on the order of \SI{100}{\mega\hertz}.}}
These signals are named: $\acrshort{TCom}_{\texttt{1A}}$, $\acrshort{TCom}_{\texttt{1B}}$, $\acrshort{TCom}_{\texttt{2A}}$, and~$\acrshort{TCom}_{\texttt{2B}}$; where the first subscript indicates the source and the second the detector. The~short \gls{rho} measurements ($\acrshort{rho}=\SI{14.9}{\milli\meter}$) are $\acrshort{TCom}_{\texttt{1A}}$ and $\acrshort{TCom}_{\texttt{2B}}$ while the long \gls{rho} measurements ($\acrshort{rho}=\SI{25.1}{\milli\meter}$) are $\acrshort{TCom}_{\texttt{2A}}$ and $\acrshort{TCom}_{\texttt{1B}}$. \par

\begin{figure}[H]
	\includegraphics{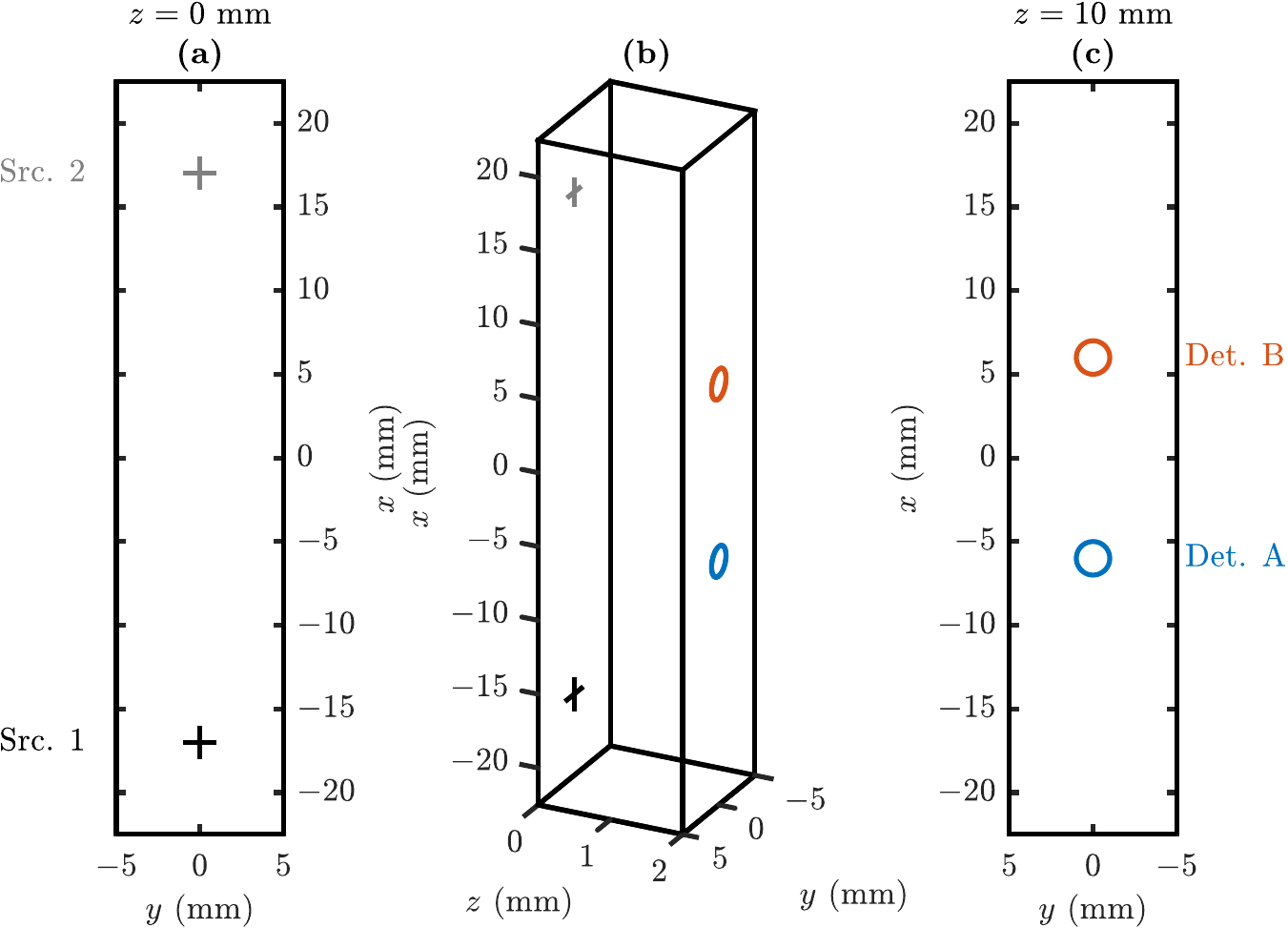}
	\caption{\hl{Schematic diagram} %MDPI: Please change the hyphen (-) indicating negative into a minus sign (−, “U+2212”), e.g., “-1” should be “−1”. We moved Figure 1 from Section 2.1 to Section 2.2 to reduce the unnecessary white blank, please confirm.
	% GB: Done and Confirmed
 of cuvette geometry measuring \SI{45x10x10}{\milli\meter}. Sources and detectors are considered at the following \acrshortpl{r}:  $\acrshort{r}_{\texttt{1}}=-17 \hat{x}~\si{\milli\meter}$, $\acrshort{r}_{\texttt{2}}=17 \hat{x}~\si{\milli\meter}$, $\acrshort{r}_{\texttt{A}}=-6 \hat{x} +10 \hat{z}~\si{\milli\meter}$, and~$\acrshort{r}_{\texttt{B}}=6 \hat{x} +10 \hat{z}~\si{\milli\meter}$. (\textbf{a}) $y-x$ plane for $z=\SI{0}{\milli\meter}$. (\textbf{b}) Transparent projected view. (\textbf{c}) $y-x$ plane for $z=\SI{10}{\milli\meter}$. {Acronyms and Symbols: \Acrfull{r}}.
		\label{fig:geo}}
\end{figure}

From these \gls{TCom} measurements, ratios between the short and long \gls{rho} measurements may be obtained. Therefore we introduce the \glspl{SRTCom} for the geometry in Figure~\ref{fig:geo} as follows:
\begin{equation}\label{equ:SRT_1AB}
	\acrshort{SRTCom}_{\text{\texttt{1AB}}}=\frac{\acrshort{TCom}_{\text{\texttt{1B}}}}{\acrshort{TCom}_{\text{\texttt{1A}}}}
\end{equation}
\begin{equation}\label{equ:SRT_2BA}
	\acrshort{SRTCom}_{\text{\texttt{2BA}}}=\frac{\acrshort{TCom}_{\text{\texttt{2A}}}}{\acrshort{TCom}_{\text{\texttt{2B}}}}
\end{equation}
\noindent and the \gls{DRTCom} as the geometric mean of the two symmetric \glspl{SRTCom}.
\begin{equation}\label{equ:DRT}
	\acrshort{DRTCom}_{\text{\texttt{1AB2}}}=\sqrt{\acrshort{SRTCom}_{\text{\texttt{1AB}}}\times\acrshort{SRTCom}_{\text{\texttt{2BA}}}}=
		\sqrt{\frac{\acrshort{TCom}_{\text{\texttt{1B}}}\acrshort{TCom}_{\text{\texttt{2A}}}}{\acrshort{TCom}_{\text{\texttt{1A}}}\acrshort{TCom}_{\text{\texttt{2B}}}}}
\end{equation}

 {This forms}   
 a similar type of measurement to \gls{DS}/\gls{SC} but replacing the concept of slope with that of ratio  \hl{(}The word slope in \acrfull{DS} is historical~\cite{sassaroliDualslopeMethodEnhanced2019,blaneyPhaseDualslopesFrequencydomain2020} as no slopes are actually considered in this work\hl{)}.
\q{R4C5}{{We acknowledge that the notation used here is verbose since, that even for \gls{DRTCom} we utilize subscripts to show all optodes used. However, for~this work we have opted to use this notation to distinguish explicitly, the~origin of the measurements. This is helpful in observing the differences in \glspl{SRTCom}, particularly in Section~\ref{sec:selfCal}. In~other work we opt to utilize a simpler notation with numbered subscripts~\cite{blaneyPhaseDualslopesFrequencydomain2020}.}} \par
We can further expand the expression for \gls{SRTCom} to consider \gls{SRTComI} and \gls{SRTComP}. For example, with \texttt{1AB} we have:
\begin{equation}
	\gls{SRTComI}_{\text{\texttt{1AB}}}=\frac{|\acrshort{TCom}_{\text{\texttt{1B}}}|}{|\acrshort{TCom}_{\text{\texttt{1A}}}|}
\end{equation}
\begin{equation}\label{equ:dP_1AB}
	\gls{SRTComP}_{\text{\texttt{1AB}}}=\angle\acrshort{TCom}_{\text{\texttt{1B}}}-\angle\acrshort{TCom}_{\text{\texttt{1A}}}
\end{equation}
\noindent and we introduce the final ratio type, the~\gls{SRTComLnI}:
\begin{equation}\label{equ:dI_1AB}
	\gls{SRTComLnI}_{\text{\texttt{1AB}}}=\ln|\acrshort{TCom}_{\text{\texttt{1B}}}|-\ln|\acrshort{TCom}_{\text{\texttt{1A}}}|
\end{equation}

 {The motivation} for utilizing the natural logarithm in this way is that it partly linearizes typical expressions for diffuse optical measurements (\gls{TCom} in this case) as a function of \gls{rho}~\cite{fantiniNoninvasiveOpticalMonitoring1999}. Additionally it shows a symmetry between \gls{SRTComLnI} and \gls{SRTComP} as they are both differences. This work focuses on utilizing these \gls{SRTComLnI} and \gls{SRTComP} ratio types in the development of the proposed~method. \par

Note that similar expressions can also be written for \gls{DRTComI}, \gls{DRTComP}, and~\gls{DRTComLnI}:
\begin{equation}
	\gls{DRTComI}_{\text{\texttt{1AB2}}}=\sqrt{\gls{SRTComI}_{\text{\texttt{1AB}}}\times\gls{SRTComI}_{\text{\texttt{2BA}}}}=\sqrt{\frac{|\acrshort{TCom}_{\text{\texttt{1B}}}||\acrshort{TCom}_{\text{\texttt{2A}}}|}{|\acrshort{TCom}_{\text{\texttt{1A}}}||\acrshort{TCom}_{\text{\texttt{2B}}}|}}
\end{equation}
\begin{equation}
	\gls{DRTComP}_{\text{\texttt{1AB2}}}=\frac{\gls{SRTComP}_{\text{\texttt{1AB}}}+\gls{SRTComP}_{\text{\texttt{2BA}}}}{2}=\frac{\angle\acrshort{TCom}_{\text{\texttt{1B}}}+\angle\acrshort{TCom}_{\text{\texttt{2A}}}-\angle\acrshort{TCom}_{\text{\texttt{1A}}}-\angle\acrshort{TCom}_{\text{\texttt{2B}}}}{2}
\end{equation}
\begin{equation}
	\begin{split}
		\gls{DRTComLnI}_{\text{\texttt{1AB2}}}=
		\frac{\gls{SRTComLnI}_{\text{\texttt{1AB}}}+\gls{SRTComLnI}_{\text{\texttt{2BA}}}}{2}=\\
		\frac{\ln|\acrshort{TCom}_{\text{\texttt{1B}}}|+\ln|\acrshort{TCom}_{\text{\texttt{2A}}}|-\ln|\acrshort{TCom}_{\text{\texttt{1A}}}|-\ln|\acrshort{TCom}_{\text{\texttt{2B}}}|}{2}
	\end{split}	
\end{equation}

 From this it can be seen that \gls{DRTComI} is a geometric mean of \glspl{SRTComI} and both \gls{DRTComP} as well as \gls{DRTComLnI} are arithmetic means of \glspl{SRTComP} and \glspl{SRTComLnI}, respectively.

For theoretical calculations, not considering optode coupling differences and medium heterogeneity, the~different \glspl{SRTCom} and \glspl{DRTCom} have the same value. This is due to the symmetry shown in Figure~\ref{fig:geo} considering a homogeneous medium. For~this reason, only the set \texttt{1AB}, and~$\acrshort{SRTCom}_{\text{\texttt{1AB}}}$, is considered for most of the results. Coupling is considered in Section~\ref{sec:selfCal}, therefore, discrepancies between the difference measurements are investigated in that section and the distinction between \gls{SRTCom} and \gls{DRTCom} becomes important~there. \par

\subsection{Analytical Box~Model}
To generate data for the cuvette geometry (Figure~\ref{fig:geo}) we utilized the following diffusion theory derived analytical expression for the \gls{TCom} \cite{kienleLightDiffusionTurbidParallelepiped2005} \hl{(}The expression used for the \acrfull{TCom} represents the measured transmittance normalized by the source power giving it units of \si{\per\square\milli\meter}\hl{)}:

\begin{equation}\label{equ:TCom}
	\begin{split}
		\acrshort{TCom}(x_{Det.},y_{Det.},z_{Det.}=L_z)=\frac{1}{4\pi}\sum_{l=-\infty}^{\infty}\sum_{m=-\infty}^{\infty}\sum_{n=-\infty}^{\infty}\Bigg[ \\
		\frac{\left(L_z-z_{1n}\right)\left(\acrshort{mueffCom}+1/r_1\right)}{r_1^2} e^{-\acrshort{mueffCom}r_1}
		-\frac{\left(L_z-z_{2n}\right)\left(\acrshort{mueffCom}+1/r_2\right)}{r_2^2} e^{-\acrshort{mueffCom}r_2} \\
		-\frac{\left(L_z-z_{1n}\right)\left(\acrshort{mueffCom}+1/r_3\right)}{r_3^2} e^{-\acrshort{mueffCom}r_3}
		+\frac{\left(L_z-z_{2n}\right)\left(\acrshort{mueffCom}+1/r_4\right)}{r_4^2} e^{-\acrshort{mueffCom}r_4} \\
		-\frac{\left(L_z-z_{1n}\right)\left(\acrshort{mueffCom}+1/r_5\right)}{r_5^2} e^{-\acrshort{mueffCom}r_5}
		+\frac{\left(L_z-z_{2n}\right)\left(\acrshort{mueffCom}+1/r_6\right)}{r_6^2} e^{-\acrshort{mueffCom}r_6} \\
		+\frac{\left(L_z-z_{1n}\right)\left(\acrshort{mueffCom}+1/r_7\right)}{r_7^2} e^{-\acrshort{mueffCom}r_7}
		-\frac{\left(L_z-z_{2n}\right)\left(\acrshort{mueffCom}+1/r_8\right)}{r_8^2} e^{-\acrshort{mueffCom}r_8}
		\Bigg]
	\end{split}
\end{equation}
\noindent where the optical properties, the~\gls{mua} and the \gls{musp}, are contained within the \gls{mueffCom}:
\begin{equation}\label{equ:mueffCom}
	\acrshort{mueffCom}=\sqrt{3\acrshort{musp}\left(\acrshort{mua}-\frac{\acrshort{omega}\acrshort{n}_i}{\acrshort{c}}i\right)}
\end{equation}
\noindent and the remaining non-spatial variables are the \gls{omega}, the~\gls{n} inside the medium ($\acrshort{n}_i$), and~the \gls{c}. \par

For spatial variables we first have the cuvette dimensions: $L_x=\SI{45}{\milli\meter}$, $L_y=\SI{10}{\milli\meter}$, and~$L_z=\SI{10}{\milli\meter}$ (Figure~\ref{fig:geo}). Next, we have the source coordinates: $x_{Src.}$ and $y_{Src.}$ (given that a pencil beam impinges on the $z=\SI{0}{\milli\meter}$ face so that an isotropic source is placed at $z_{Iso.-Src.}=1/\acrshort{musp}$); as well as the detector coordinates: $x_{Det.}$ and $y_{Det.}$ (given that the detector is placed on the $z=L_z$ face, thus $z_{Det.}=L_z$). Using these variables and remembering the sum indexes from Equation (\ref{equ:TCom}) we can now write the various point source positions (infinite of both positive and negative as per the indexing variables $l$, $m$, and~$n$) \cite{kienleLightDiffusionTurbidParallelepiped2005}:
\begin{equation}
	x_{1l}=2lL_x+4lh+x_{Src.}
\end{equation}
\begin{equation}
	x_{2l}=\left(2l-1\right)L_x+\left(4l-2\right)h-x_{Src.}
\end{equation}
\begin{equation}
	y_{1m}=2mL_y+4mh+y_{Src.}
\end{equation}
\begin{equation}
	y_{2m}=\left(2m-1\right)L_y+\left(4m-2\right)h-y_{Src.}
\end{equation}
\begin{equation}
	z_{1n}=2nL_z+4nh+1/\acrshort{musp}
\end{equation}
\begin{equation}
	z_{2n}=2nL_z+\left(4n-2\right)h-1/\acrshort{musp}
\end{equation}
\noindent where $h$ is the distance between the extrapolated boundary and the actual box boundary:
\begin{equation}\label{equ:exBound}
	h=\frac{2a(\acrshort{n}_r)}{3\acrshort{musp}}
\end{equation}
\noindent $a$ is the \gls{n} mismatch parameter~\cite{continiPhotonMigrationTurbid1997,martelliPhotonMigrationTurbid1997} which is a function of the relative \gls{n} mismatch ($\acrshort{n}_r=\acrshort{n}_i/\acrshort{n}_o$, where $\acrshort{n}_o$ is the \gls{n} outside). Finally, using these positions, we can define the distances to the point sources:
\begin{equation}
	r_1=\sqrt{\left(x_{Det.}-x_{1l}\right)^2+\left(y_{Det.}-y_{1m}\right)^2+\left(z_{Det.}-z_{1n}\right)^2}
\end{equation}
\begin{equation}
	r_2=\sqrt{\left(x_{Det.}-x_{1l}\right)^2+\left(y_{Det.}-y_{1m}\right)^2+\left(z_{Det.}-z_{2n}\right)^2}
\end{equation}
\begin{equation}
	r_3=\sqrt{\left(x_{Det.}-x_{1l}\right)^2+\left(y_{Det.}-y_{2m}\right)^2+\left(z_{Det.}-z_{1n}\right)^2}
\end{equation}
\begin{equation}
	r_4=\sqrt{\left(x_{Det.}-x_{1l}\right)^2+\left(y_{Det.}-y_{2m}\right)^2+\left(z_{Det.}-z_{2n}\right)^2}
\end{equation}
\begin{equation}
	r_5=\sqrt{\left(x_{Det.}-x_{2l}\right)^2+\left(y_{Det.}-y_{1m}\right)^2+\left(z_{Det.}-z_{1n}\right)^2}
\end{equation}
\begin{equation}
	r_6=\sqrt{\left(x_{Det.}-x_{2l}\right)^2+\left(y_{Det.}-y_{1m}\right)^2+\left(z_{Det.}-z_{2n}\right)^2}
\end{equation}
\begin{equation}
	r_7=\sqrt{\left(x_{Det.}-x_{2l}\right)^2+\left(y_{Det.}-y_{2m}\right)^2+\left(z_{Det.}-z_{1n}\right)^2}
\end{equation}
\begin{equation}
	r_8=\sqrt{\left(x_{Det.}-x_{2l}\right)^2+\left(y_{Det.}-y_{2m}\right)^2+\left(z_{Det.}-z_{2n}\right)^2}
\end{equation}
\par

\q{R1C3a}{{This diffusion theory derived expression (Equation~(\ref{equ:TCom})) was previously presented and validated against Monte Carlo in~\cite{kienleLightDiffusionTurbidParallelepiped2005}. The~validity of such an expression is dependent on the distance from the source, the~value of \gls{musp}, and~the ratio between \gls{mua} and \gls{musp}.}}
\q{R2C2}{{Only locations far enough from the source for scattering to be considered isotropic may be considered, \gls{musp} must be large enough for isotropic scattering to dominate within the medium, and~\gls{mua} must be much less than \gls{musp}. Given that we consider measurements on the opposite side of a \SI{10}{\milli\meter} thick cuvette the first and second conditions are met for \gls{musp} values on the order of \SI{1}{\per\milli\meter} (cuvette thickness of \num{10} isotropic scattering mean free paths). Finally, the~last condition may be met by considering values of \gls{mua} on the order of \num{100} times smaller than \gls{musp}. For~this work, we choose values which encompass the edge of validity of diffusion theory with \gls{mua} from \SIrange{0}{0.05}{\per\milli\meter} and \gls{musp} from \SIrange{0.5}{5}{\per\milli\meter} for demonstration purposes.}}
\q{R1C3b}{{Most focus is on the values of \SI{0.01}{\per\milli\meter} for \gls{mua} and \SI{1}{\per\milli\meter} for \gls{musp}, for~which our independent (aside from what is presented in~\cite{kienleLightDiffusionTurbidParallelepiped2005}) validation against Monte Carlo~\cite{fangMonteCarloSimulation2009} found \SI{<5}{\percent} discrepancy for $|\acrshort{TCom}|$ and \SI{0.005}{\radian} for $\angle\acrshort{TCom}$ over the face of the cuvette opposing the source. For~the purposes of this work, a~feasibility study of the proposed method, we believe this diffusion theory model is appropriate.}} \par

\q{R4C2}{{To provide some physical intuition regarding the \gls{FD} portion of this model, we can discuss the wavelength of the photon density waves ($\lambda_{PDW}$), which can be approximated as~\cite{bigioQuantitativeBiomedicalOptics2016}:
\begin{equation}
	\lambda_{PDW}=\frac{2\pi}{\sqrt{\frac{3}{2} \acrshort{mua}\acrshort{musp}\left(\sqrt{1+\left(\frac{\acrshort{omega}\acrshort{n}}{\acrshort{c}\acrshort{mua}}\right)^2}-1\right)}}
\end{equation}

 This leads to a $\lambda_{PDW}$ of about \SI{160}{\milli\meter} using values of \SI{0.01}{\per\milli\meter} for \gls{mua}, \SI{1}{\per\milli\meter} for \gls{musp}, $2\pi\times\SI{100e6}{\radian\per\second}$ for \gls{omega}, \num{1} for \gls{n}, and~\SI{2.99792458e11}{\milli\meter\per\second} for \gls{c}. This is larger then the cuvette, but~not so large that it dwarfs the cuvette scale completely. This indicates that phase measurements are reasonable for this volume, given that the phase will not wrap ($\lambda_{PDW}$ not too short) but will still change considerably throughout the size of the cuvette ($\lambda_{PDW}$ not too long).}} \par

\begin{figure}[H]
	\begin{center}
		\hspace{-25pt}
\includegraphics[scale=0.94]{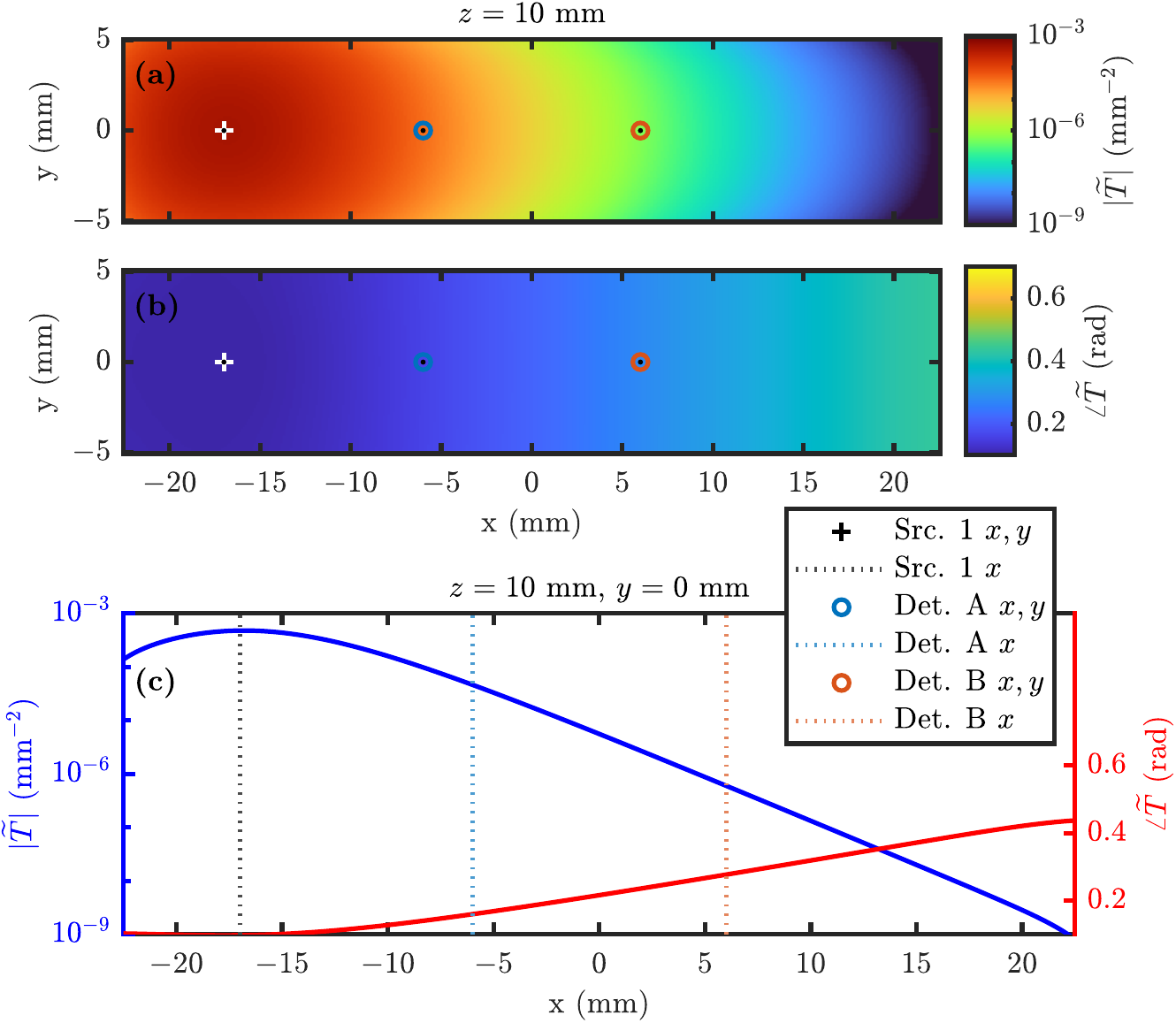}
	\end{center}
	\caption{\hl{Example of the} %MDPI: Please change the hyphen (-) indicating negative into a minus sign (−, “U+2212”), e.g., “-1” should be “−1”. We moved Figure 1 from Section 2.3 to Section 2.4 to reduce the unnecessary white blank, please confirm.
	% GB: Done and Confirmed
 implementation of the diffusion theory derived expression for \acrshort{TCom} in \linebreak Equation~(\ref{equ:TCom}) showing \acrshort{TCom}  \hl{(}The expression used for the \acrfull{TCom} represents the measured transmittance normalized by the source power giving it units of \si{\per\square\milli\meter}\hl{)} on the cuvette face opposing the source ($z=L_z=\SI{10}{\milli\meter}$) considering the geometry in Figure~\ref{fig:geo} and source~\texttt{1}. For~this simulation the \acrshort{mua} was \SI{0.01}{\per\milli\meter}, the~\acrshort{musp} \SI{1}{\per\milli\meter}, \acrshort{n} inside \num{1.3}, \acrshort{n} outside \num{1}, \acrshort{omega} $2\pi\SI{100e6}{\radian\per\second}$, and~the cuvette measured \SI{45x10x10}{\milli\meter}. The~source ($x-y$ position is shown as cross; $x$ position as black dotted line) was placed at $\acrshort{r}_{\text{Src.\texttt{1}}}=-17 \hat{x}~\si{\milli\meter}$. Detector positions which are considered in the following work are shown as circles (for $x-y$ position) or dotted lines (for $x$ position) (\textbf{a}) \acrshort{TCom} amplitude ($|\acrshort{TCom}|$) on the $x-y$ plane at $z=L_z=\SI{10}{\milli\meter}$. (\textbf{b}) \acrshort{TCom} phase ($\angle\acrshort{TCom}$) on the $x-y$ plane at $z=L_z=\SI{10}{\milli\meter}$. (\textbf{c}) $|\acrshort{TCom}|$ and $\angle\acrshort{TCom}$ along the $x$ direction for $z=L_z=\SI{10}{\milli\meter}$ and $y=\SI{0}{\milli\meter}$. Acronyms and Symbols: \Acrfull{TCom}, \acrfull{mua}, \acrfull{musp}, \acrfull{n}, \acrfull{omega}, and~\acrfull{r}.
		\label{fig:exDT}}
\end{figure}

To demonstrate the implementation of this expression for \gls{TCom} (Equation~(\ref{equ:TCom})) we show a map of the amplitude ($|\acrshort{TCom}|$) and phase ($\angle\acrshort{TCom}$) on the $z=L_z=\SI{10}{\milli\meter}$ face (opposing the source; Figure~\ref{fig:geo}) for $\acrshort{mua}=\SI{0.01}{\per\milli\meter}$ and $\acrshort{musp}=\SI{1}{\per\milli\meter}$ in Figure~\ref{fig:exDT} (considering \mbox{source \texttt{1}}). This shows the spatial continuum of \gls{TCom} which can be simulated with diffusion theory. The~positions of the source (\texttt{1}) and detectors (\texttt{A} \& \texttt{B}) are also indicated in Figure~\ref{fig:exDT} to show the positions which will be considered throughout this work. For~computation based on Equation~(\ref{equ:TCom}), $l$, $m$, and~$n$ were each summed from \numrange{-3}{3}, and~inclusion of more terms was found to not significantly impact the~results. \par

%\unskip

\subsection{Optical Properties~Fit}\label{sec:fitCost}
{The end goal of this work is to develop a method capable of measuring the absolute \gls{mua} and \gls{musp} in the geometry of Figure~\ref{fig:geo} using \gls{FD}.} {With this in mind, we define a cost ($\chi^2$) function which can be minimized by varying \gls{mua} and \gls{musp}, thus creating a fit for \gls{mua} and \gls{musp} \hl{(}We acknowledge that the cost ($\chi^2$) function is dependent on parameters beyond \acrfull{mua} and \acrfull{musp} such as \acrfull{n}, and~investigate this in further sections of this work\hl{)}: } 

\begin{equation}\label{equ:chi2}
	\begin{split}
		\chi^2(\acrshort{mua}, \acrshort{musp})=\kappa\left(\frac{\left[\acrshort{DRTComLnI}\right]_{meas}-\left[\acrshort{DRTComLnI}\right]_{theo}(\acrshort{mua}, \acrshort{musp})}{\sigma_{\acrshort{DRTComLnI}}}\right)^2\\
		+\left(\frac{\left[\acrshort{DRTComP}\right]_{meas}-\left[\acrshort{DRTComP}\right]_{theo}(\acrshort{mua}, \acrshort{musp})}{\sigma_{\acrshort{DRTComP}}}\right)^2
	\end{split}
\end{equation}
where the $meas$ subscript represents the measured difference \hl{(}In this work, the measurement is simulated using Equation~(\ref{equ:TCom}), and~noise may be added depending on the purpose\hl{)} and the $theo$ subscript represents the value retrieved from Equation~(\ref{equ:TCom}) considering a particular \gls{mua} and \gls{musp} \hl{(}Again we note that the \acrfull{DRTCom} and a \acrfull{SRTCom} are the same when not considering optode coupling (shown in Section~\ref{sec:selfCal})\hl{)}. \par
Three further variables are introduced in Equation~(\ref{equ:chi2}), which we define below. First is the \gls{DRTComLnI} scaling coefficient ($\kappa$) which is discussed further in Section~\ref{sec:kappa}. Second and third are the uncertainties of \gls{DRTComLnI} ($\sigma_{\acrshort{DRTComLnI}}$) and \gls{DRTComP} ($\sigma_{\acrshort{DRTComP}}$) which are expressed based on 1\textsuperscript{st} order error propagation as:
\begin{equation}
	\sigma_{\acrshort{SRTComLnI}}=\sqrt{\left(\frac{\sigma_{|\acrshort{TCom}|_{\text{long}}}}{|\acrshort{TCom}|_{\text{long}}}\right)^2+\left(\frac{\sigma_{|\acrshort{TCom}|_{\text{short}}}}{|\acrshort{TCom}|_{\text{short}}}\right)^2}
\end{equation}
\begin{equation}
	\sigma_{\acrshort{SRTComP}}=\sqrt{\left(\sigma_{\angle\acrshort{TCom}_{\text{long}}}\right)^2+\left(\sigma_{\angle\acrshort{TCom}_{\text{short}}}\right)^2}
\end{equation}
\noindent and
\begin{equation}
	\sigma_{\acrshort{DRTComLnI}}=\frac{\sigma_{\acrshort{SRTComLnI}}}{\sqrt{2}}
\end{equation}
\begin{equation}
	\sigma_{\acrshort{DRTComP}}=\frac{\sigma_{\acrshort{SRTComP}}}{\sqrt{2}}
\end{equation}
\noindent where, $\sigma_{|\acrshort{TCom}|}$ is the uncertainty in $|\acrshort{TCom}|$ and $\sigma_{\angle\acrshort{TCom}}$ is the uncertainty in $\angle\acrshort{TCom}$, and~assuming that the uncertainties in the two \glspl{SRTComLnI} ($\sigma_{\acrshort{SRTComLnI}}$) and the two \glspl{SRTComP} ($\sigma_{\acrshort{SRTComP}}$) are each the same. For~this work we set $\sigma_{|\acrshort{TCom}|}/|\acrshort{TCom}|=\num{0.001}$ and $\sigma_{\angle\acrshort{TCom}}=\SI{1.7}{\milli\radian}=\ang{0.1}$ which would be typical for a \gls{FD} \gls{NIRS} instrument such as the \gls{ISSv2}. \par

%%%%%%%%%%%%%%%%%%%%%%%%%%%%%%%%%%%%%%%%%%
\section{Results}
\subsection{Investigation of Difference~Measurements}
\subsubsection{Variation over Optical~Properties}
\begin{figure}[H]
	\includegraphics{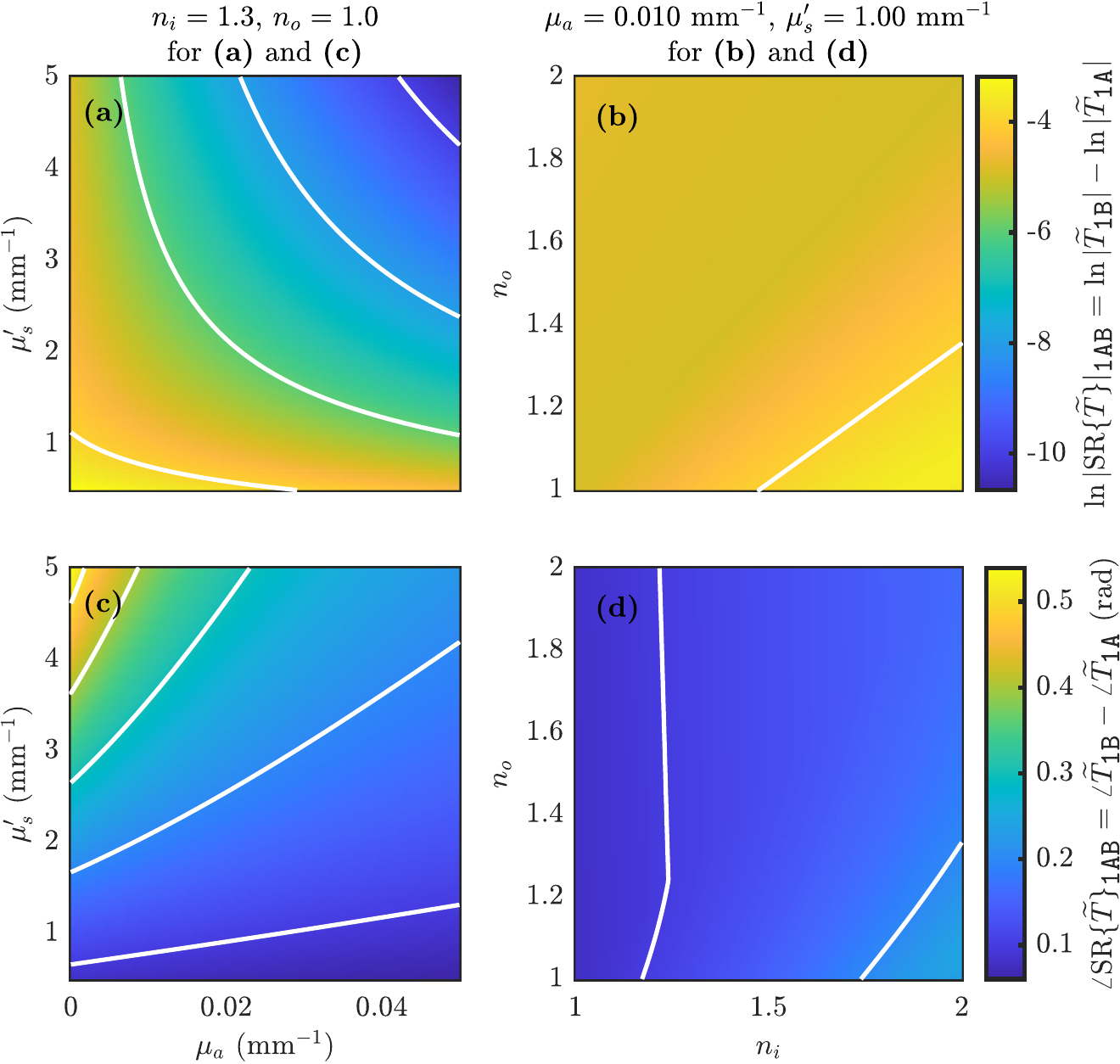}
	\caption{\hl{How the measurements} %MDPI: We moved Figure 1 from Section 3.1.1 to Section 3.1.2 to reduce the unnecessary white blank, please confirm.
	% GB: Confirmed
	of the \acrshort{SRTComLnI} and \acrshort{SRTComP} (Equations~(\ref{equ:dI_1AB}) and (\ref{equ:dP_1AB})) are effected by optical parameters, namely the \acrshort{mua}, the~\acrshort{musp}, the~\acrshort{n} inside ($\acrshort{n}_i$) and the \acrshort{n} outside ($\acrshort{n}_o$). Simulation geometry and parameters not explicitly shown here are stated in detail in Figures~\ref{fig:geo} and \ref{fig:exDT}. (\textbf{a})~\acrshort{SRTComLnI} versus \acrshort{mua} and \acrshort{musp}. (\textbf{b}) \acrshort{SRTComLnI} versus $\acrshort{n}_i$ and $\acrshort{n}_o$. (\textbf{c}) \acrshort{SRTComP} versus \acrshort{mua} and \acrshort{musp}. (\textbf{d}) \acrshort{SRTComP} versus $\acrshort{n}_i$ and $\acrshort{n}_o$. 
	Note: {(\textbf{a},\textbf{b})} have the same iso-line and color-map values/scales; as do (\textbf{c},\textbf{d}). 
	Acronyms and Symbols: \Acrfull{TCom}, \acrfull{SRTCom}, \acrfull{SRTComI}, \acrfull{SRTComLnI}, \acrfull{SRTComP}, \acrfull{mua}, \acrfull{musp}, and~\acrfull{n}.
	\label{fig:diffGrd}}
\end{figure}

The chief measurements which we consider are \gls{SRTComLnI} and \gls{SRTComP} (or \gls{DRTComLnI} and \gls{DRTComP} considering coupling; \hl{(}Again we note that the \acrfull{DRTCom} and a \acrfull{SRTCom} are the same when not considering optode coupling (shown in Section~\ref{sec:selfCal})\hl{)}). Figure~\ref{fig:diffGrd} shows these measurements (Equations~(\ref{equ:dP_1AB}) and (\ref{equ:dI_1AB})) over a large range of optical properties, specifically \gls{mua} and \gls{musp} (Figure~\ref{fig:diffGrd}{a,c}) or $\acrshort{n}_i$ and $\acrshort{n}_o$ (Figure~\ref{fig:diffGrd}{b,d}). \par

Since the intention is to convert these measurements of \gls{SRTComLnI} and \gls{SRTComP} to \gls{mua} and \gls{musp}, the~desire is for the measurements to vary significantly more as \gls{mua} and \gls{musp} are varied as compared to varying $\acrshort{n}_i$ and $\acrshort{n}_o$. The~iso-lines (white lines) in Figure~\ref{fig:diffGrd}{a,b} consider the same values (the same is true for Figure~\ref{fig:diffGrd}{c,d}). {From this we see that varying \gls{mua} and \gls{musp} varies \gls{SRTComLnI} across \num{4} more iso-lines than varying $\acrshort{n}_i$ and $\acrshort{n}_o$ (and about \num{2} times more for \gls{SRTComP}).} This suggests promise in the goal of retrieving \gls{mua} and \gls{musp}. \par

To recover \gls{mua} and \gls{musp} from \gls{SRTComLnI} and \gls{SRTComP} we must also have significantly different information in \gls{SRTComLnI} and \gls{SRTComP} so that the recovered variables (\gls{mua} and \gls{musp}) have a unique solution and little cross-talk. There is also promise along these lines as the iso-lines in Figure~\ref{fig:diffGrd}a versus Figure~\ref{fig:diffGrd}c are qualitatively orthogonal. This suggests a fit to \gls{mua} and \gls{musp} from \gls{SRTComLnI} and \gls{SRTComP} should be possible. This is further investigated in Section~\ref{sec:fit}. \par

One final insight that can be drawn from Figure~\ref{fig:diffGrd} is the effect of $\acrshort{n}_r$ which is constant along diagonal lines with positive slopes in Figure~\ref{fig:diffGrd}{b,d}. From~this we see that \gls{SRTComLnI} is little effected by $\acrshort{n}_r$. Further, in~the upper left portion of the plots where $\acrshort{n}_r<1$ \gls{SRTComP} is only significantly effected by $\acrshort{n}_i$. Therefore we may be able to optimize the design of the cuvette boundary to reduce cross-talk with the \glspl{n} which is discussed further in Section~\ref{sec:dis}. \par

%\unskip

\subsubsection{Optode Coupling and~Auto-Calibration}\label{sec:selfCal}
As has been stated above including in \hl{(}Again we note that the \acrfull{DRTCom} and a \acrfull{SRTCom} are the same when not considering optode coupling (shown in Section~\ref{sec:selfCal})\hl{)}, \gls{SRTCom} \& \gls{DRTCom} (as well as \gls{SRTComI} \& \gls{DRTComI}, \gls{SRTComP} \& \gls{DRTComP}, and~\gls{SRTComLnI} \& \gls{DRTComLnI}) are equivalent when optode coupling is not considered. {For this reason, other sections of this manuscript are not careful to distinguish between them as theoretical calculations are being carried out and coupling is not a consideration.} However, in~this section we show the effect of optode coupling and the auto-calibration of the \gls{DRTCom} which is inherited/inspired by the \gls{SC} method~\cite{hueberNewOpticalProbe1999}. \par
To do this, first we define a \gls{CCom} for each optode: $\acrshort{CCom}_{\texttt{1}}$, $\acrshort{CCom}_{\texttt{2}}$, $\acrshort{CCom}_{\texttt{A}}$, and~$\acrshort{CCom}_{\texttt{B}}$. Physically, $|\acrshort{CCom}|$ represents a multiplicative factor (attenuation or amplification) on the amplitude of \gls{TCom} and $\angle\acrshort{CCom}$ represents a phase shift on the phase of \gls{TCom}. \Glspl{CCom} applied to sources (number subscripts) have units of \si{\milli\watt} since their amplitude also includes source power; while \glspl{CCom} for detectors (letter subscripts) are unit-less. Therefore, adding $coup$ subscripts to our measurements when they are confounded by coupling (opposed to the theoretical value without the $coup$ subscript) we have the following signals considering coupling:
\begin{equation}\label{equ:T1A_coup}
	\acrshort{TCom}_{\texttt{1A},coup}=\acrshort{CCom}_{\texttt{1}}\acrshort{CCom}_{\texttt{A}}\acrshort{TCom}_{\texttt{1A}}
\end{equation}
\begin{equation}
	\acrshort{TCom}_{\texttt{1B},coup}=\acrshort{CCom}_{\texttt{1}}\acrshort{CCom}_{\texttt{B}}\acrshort{TCom}_{\texttt{1B}}
\end{equation}
\begin{equation}
	\acrshort{TCom}_{\texttt{2A},coup}=\acrshort{CCom}_{\texttt{2}}\acrshort{CCom}_{\texttt{A}}\acrshort{TCom}_{\texttt{2A}}
\end{equation}
\begin{equation}\label{equ:T2B_coup}
	\acrshort{TCom}_{\texttt{2B},coup}=\acrshort{CCom}_{\texttt{2}}\acrshort{CCom}_{\texttt{B}}\acrshort{TCom}_{\texttt{2B}}
\end{equation}
\noindent Now, let us revisit Equations~(\ref{equ:SRT_1AB})--(\ref{equ:DRT}) but with optode coupling considered:
\begin{equation}
	\acrshort{SRTCom}_{\text{\texttt{1AB}},coup}=\frac{\acrshort{CCom}_{\texttt{1}}\acrshort{CCom}_{\texttt{B}}\acrshort{TCom}_{\text{\texttt{1B}}}}
		{\acrshort{CCom}_{\texttt{1}}\acrshort{CCom}_{\texttt{A}}\acrshort{TCom}_{\text{\texttt{1A}}}}=
		\frac{\acrshort{CCom}_{\texttt{B}}\acrshort{TCom}_{\text{\texttt{1B}}}}
		{\acrshort{CCom}_{\texttt{A}}\acrshort{TCom}_{\text{\texttt{1A}}}}=
		\frac{\acrshort{CCom}_{\texttt{B}}}{\acrshort{CCom}_{\texttt{A}}}\acrshort{SRTCom}_{\text{\texttt{1AB}}}
\end{equation}
\begin{equation}
	\acrshort{SRTCom}_{\text{\texttt{2BA}},coup}=\frac{\acrshort{CCom}_{\texttt{2}}\acrshort{CCom}_{\texttt{A}}\acrshort{TCom}_{\text{\texttt{2A}}}}
		{\acrshort{CCom}_{\texttt{2}}\acrshort{CCom}_{\texttt{B}}\acrshort{TCom}_{\text{\texttt{2B}}}}=
		\frac{\acrshort{CCom}_{\texttt{A}}\acrshort{TCom}_{\text{\texttt{2A}}}}
		{\acrshort{CCom}_{\texttt{B}}\acrshort{TCom}_{\text{\texttt{2B}}}}=
		\frac{\acrshort{CCom}_{\texttt{A}}}{\acrshort{CCom}_{\texttt{B}}}\acrshort{SRTCom}_{\text{\texttt{2BA}}}
\end{equation}
\begin{equation}
	\acrshort{DRTCom}_{\text{\texttt{1AB2}},coup}=
		\sqrt{\frac{\acrshort{CCom}_{\texttt{B}}\acrshort{TCom}_{\text{\texttt{1B}}}\acrshort{CCom}_{\texttt{A}}\acrshort{TCom}_{\text{\texttt{2A}}}}{\acrshort{CCom}_{\texttt{A}}\acrshort{TCom}_{\text{\texttt{1A}}}\acrshort{CCom}_{\texttt{B}}\acrshort{TCom}_{\text{\texttt{2B}}}}}=
		\sqrt{\frac{\acrshort{TCom}_{\text{\texttt{1B}}}\acrshort{TCom}_{\text{\texttt{2A}}}}{\acrshort{TCom}_{\text{\texttt{1A}}}\acrshort{TCom}_{\text{\texttt{2B}}}}}=
		\acrshort{DRTCom}_{\text{\texttt{1AB2}}}
\end{equation}
\noindent showing that the measured \gls{DRTCom} is the same as the theoretical values regardless of the various optode couplings \gls{CCom}. Notice that the same follows for \gls{DRTComI}, \gls{DRTComP}, and~\gls{DRTComLnI}. \par

\begin{figure}[H]
	\includegraphics{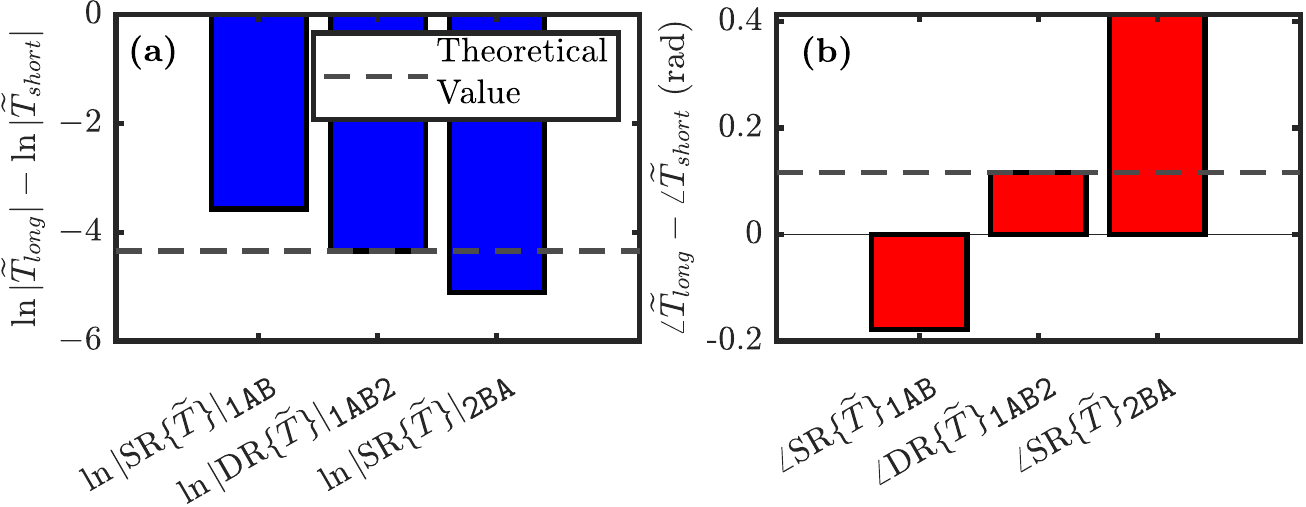}
	\caption{\hl{Demonstration of the} %MDPI: Please change the hyphen (-) into a minus sign (−, “U+2212”), e.g., “-1” should be “−1”.
	% GB: Done
	cancellation of coupling factors when considering the \acrshort{DRTComLnI} and \acrshort{DRTComP} measurements. For~this simulation, random \acrshortpl{CCom} were generated for each optode and Equations~(\ref{equ:T1A_coup})--(\ref{equ:T2B_coup}) implemented. All other simulation parameters are the same as \mbox{Figures~\ref{fig:geo} and \ref{fig:exDT}}. The~expected theoretical value for the measured differences is shown as a dashed line. (\textbf{a}) The two symmetric \acrshort{SRTComLnI} measurements and the \acrshort{DRTComLnI} measurement. (\textbf{b}) The two symmetric \acrshort{SRTComP} measurements and the \acrshort{DRTComP} measurement. 
	Acronyms and Symbols: \Acrfull{TCom}, \acrfull{SRTCom}, \acrfull{SRTComI}, \acrfull{SRTComLnI}, \acrfull{SRTComP}, \acrfull{DRTCom}, \acrfull{DRTComI}, \acrfull{DRTComLnI}, \acrfull{DRTComP}, and~\acrfull{CCom}.
	\label{fig:Tcal}}
\end{figure}

This property of auto-calibration (coming from the \gls{SC} method~\cite{hueberNewOpticalProbe1999}) is demonstrated in Figure~\ref{fig:Tcal}. However, in~this case, unlike \gls{SC}, the~symmetry requirements are not as strict since ratios instead of slopes are used as the measurement. In~this case a random \gls{CCom} was applied for each optode and the difference measurements both averaged and not were simulated. From~Figure~\ref{fig:Tcal} one can see that the \gls{DRTComLnI} and \gls{DRTComP} measurements are the same as the theoretical values. This is significant since it shows that the proposed measurement method would be insensitive to optode coupling, and~further, optode coupling would not effect the recovered \gls{mua} and \gls{musp}. Therefore, the~instrument would not need to be calibrated in terms of coupling, reducing possible systematic errors and making the method simpler to~implement. \par

\unskip

\subsection{Development of Fit for Absolute Optical~Properties}\label{sec:fit}
\unskip
\subsubsection{Optimization of Cost Space~Shape}\label{sec:kappa}
In order to fit for the absolute optical properties \gls{mua} and \gls{musp} we consider the $\chi^2$ function in Equation~(\ref{equ:chi2}). This function contains the scaling parameter $\kappa$ which balances the scale of \gls{SRTComLnI} versus \gls{SRTComP}. The~intention of such a parameter is to modify the $\chi^2$ space to be as circular as possible. This circularity can be quantitatively defined by considering iso-lines in cost space and their perimeter ($P$) as well as area ($A$). A~circle has the minimum ratio of $P$ to $A$ of all \gls{2D} shapes. Therefore, the~dimensionless metric $P^2/A$ was minimized by varying $\kappa$ (note that $P^2/A$ has a minimum theoretical value of $4\pi$, for~a circle) \cite{yangSpatiallyenhancedTimedomainNIRS2019}. \par

\begin{figure}[H]
	\includegraphics{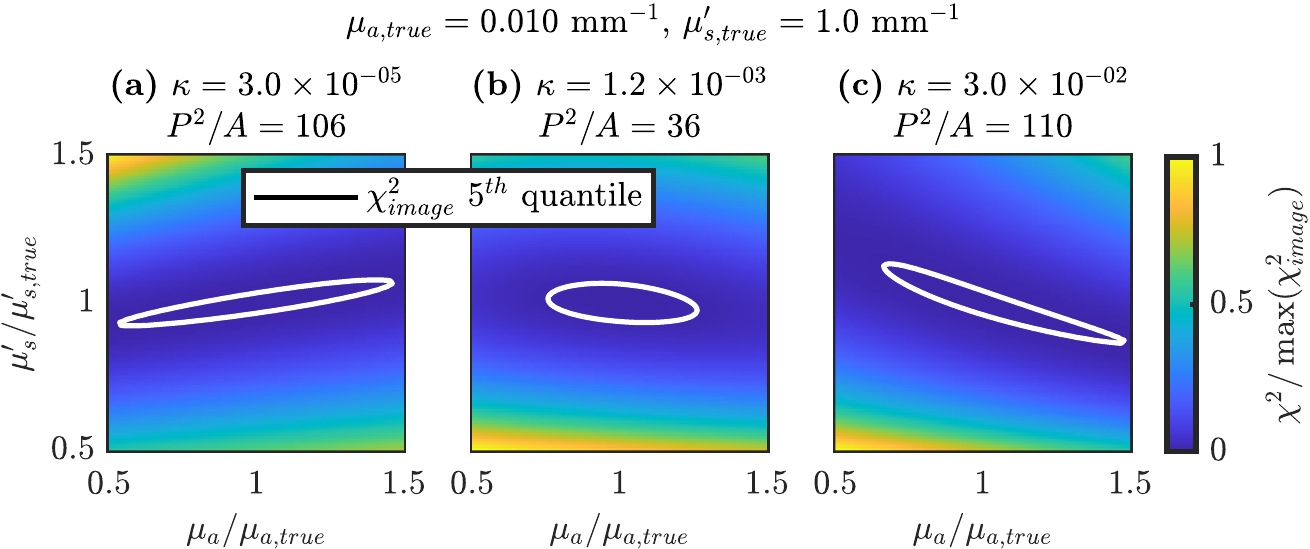}
	\caption{Three examples of the $\chi^2$ (Equation~(\ref{equ:chi2})) space shape for different $\kappa$s. The~shape of cost space is determined by the shape of an iso-line and its ratio of $P$ squared divided by $A$ ($P^2/A$) which has a minimum possible value of $4\pi$ (in the case of a circle). Iso-lines are the 5th quantile of all $\chi^2$ values in each image and are meant to represent the overall shape of $\chi^2$ space. For~all axes, $\chi^2$, the~\acrshort{mua}, and~the \acrshort{musp} are normalized. \hl{(\textbf{a}) $\kappa=\num{3e-5}$ and $P^2/A=106$}. %MDPI: it is repeated with the explanation in the picture, please confirm.
	% GB: We prefer to keep it for clarity, but it may be removed if nessisary.
 (\textbf{b}) Optimal value of $\kappa$ for the $true$ optical properties used here found by minimizing $P^2/A$, resulting in $\kappa=\num{1.2e-3}$ and $P^2/A=\num{36}$. \hl{(\textbf{c}) $\kappa=\num{3e-2}$ and $P^2/A=110$}. 
		Acronyms and Symbols: Cost ($\chi^2$), scaling factor ($\kappa$), Perimeter ($P$), Area ($A$), \acrfull{mua}, and~\acrfull{musp}.
		\label{fig:kappas}}
\end{figure}

The effect of the $\kappa$ value on $\chi^2$ space shape is shown in Figure~\ref{fig:kappas} using the same parameters as Figures~\ref{fig:geo} and \ref{fig:exDT} (where the optical properties are the $true$ values). Figure~\ref{fig:kappas}b shows the optimal $\kappa$ of \num{1.2e-3}, which is the case where $P^2/A$ was minimized. The~resulting $P^2/A$, for~this optimal $\kappa$ was \num{36} about \num{3} times worse than the value of $4\pi\approx\num{12.6}$ for a ideal circular cost space. This can be seen by how oblique the $\chi^2$ iso-lines are in the \gls{mua} direction suggesting a higher relative uncertainty in \gls{mua}. We investigate this further in Section~\ref{sec:confounds}. Figure~\ref{fig:kappas}{a,c} show the effect of favoring either the \gls{SRTComP} or \gls{SRTComLnI} term in the $\chi^2$ expression (Equation~(\ref{equ:chi2})). In~either case \gls{mua} and \gls{musp} are correlated and the space is spread more in \gls{mua}. However, when \gls{SRTComLnI} is favored (Figure~\ref{fig:kappas}c) \gls{mua} and \gls{musp} have a negative correlation while the correlation is positive when \gls{SRTComP} is favored (Figure~\ref{fig:kappas}a). \par

Note that the optimal $\kappa=\num{1.2e-3}$ was found for the \gls{mua} of \SI{0.01}{\milli\meter} and \gls{musp} of \SI{1}{\milli\meter} and a different optimal $\kappa$s may be found elsewhere for different $true$ \gls{mua} and \gls{musp}. Despite this we have opted to utilize this one $\kappa$ value for the rest of this work to reduce computation time, in~the future a map of optimal $\kappa$ could be found as a function of \gls{mua} and \gls{musp}. \par

\unskip

\subsubsection{Cost Space Shape for Various Optical~Properties}
\begin{figure}[H]
	\includegraphics[scale=0.97]{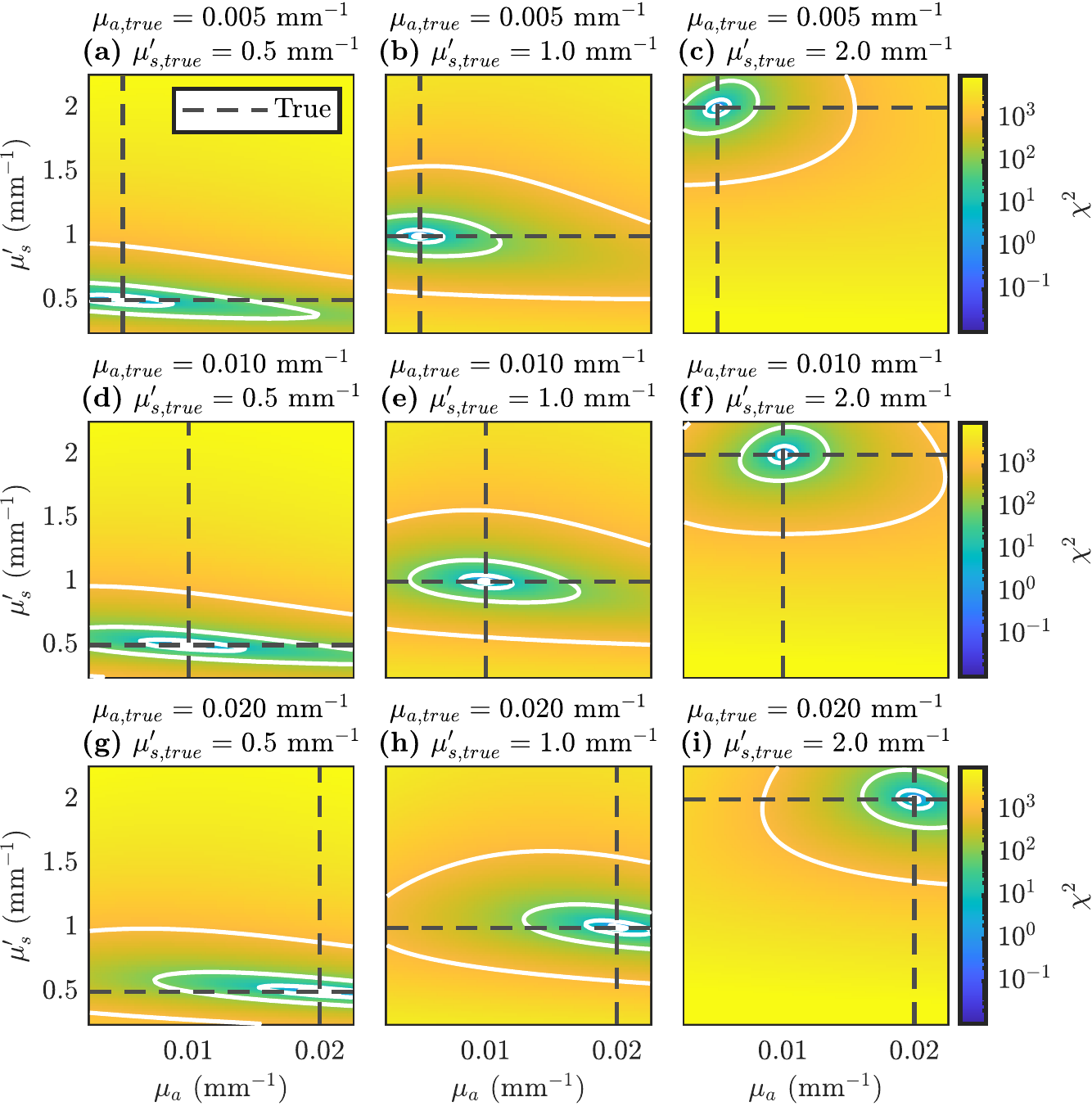}
	\caption{\num{9} examples of $\chi^2$ (Equation~(\ref{equ:chi2}); $\kappa=\num{1.2e-3}$) space for different sets of the $true$ \acrshort{mua} and \acrshort{musp}. 
	\hl{(\textbf{a})} 
 $\acrshort{mua}_{,true}=\SI{0.005}{\per\milli\meter}$ and $\acrshort{musp}_{,true}=\SI{0.5}{\per\milli\meter}$. \hl{(\textbf{b})} $\acrshort{mua}_{,true}=\SI{0.005}{\per\milli\meter}$ and $\acrshort{musp}_{,true}=\SI{1.0}{\per\milli\meter}$.
	\hl{(\textbf{c})} $\acrshort{mua}_{,true}=\SI{0.005}{\per\milli\meter}$ and $\acrshort{musp}_{,true}=\SI{2.0}{\per\milli\meter}$. \hl{(\textbf{d})} $\acrshort{mua}_{,true}=\SI{0.010}{\per\milli\meter}$ and $\acrshort{musp}_{,true}=\SI{0.5}{\per\milli\meter}$.
	\hl{(\textbf{e})} $\acrshort{mua}_{,true}=\SI{0.010}{\per\milli\meter}$ and $\acrshort{musp}_{,true}=\SI{1.0}{\per\milli\meter}$. \hl{(\textbf{f})} $\acrshort{mua}_{,true}=\SI{0.010}{\per\milli\meter}$ and $\acrshort{musp}_{,true}=\SI{2.0}{\per\milli\meter}$.
	\hl{(\textbf{g})} $\acrshort{mua}_{,true}=\SI{0.020}{\per\milli\meter}$ and $\acrshort{musp}_{,true}=\SI{0.5}{\per\milli\meter}$. \hl{(\textbf{h})} $\acrshort{mua}_{,true}=\SI{0.020}{\per\milli\meter}$ and $\acrshort{musp}_{,true}=\SI{1.0}{\per\milli\meter}$.
	\hl{(\textbf{i})} $\acrshort{mua}_{,true}=\SI{0.020}{\per\milli\meter}$ and $\acrshort{musp}_{,true}=\SI{2.0}{\per\milli\meter}$. 
		Acronyms and Symbols: Cost ($\chi^2$), scaling factor ($\kappa$), \acrfull{mua}, and~\acrfull{musp}. %MDPI: the explanation of the subfigures are repeated with the explanation in the picture, please confirm if they can be deleted.
		% GB: We prefer to keep it for clarity, but it may be removed if nessisary.
		\label{fig:costSpace}}
\end{figure}

Now that the entire cost ($\chi^2$; Equation~(\ref{equ:chi2})) function including $\kappa=\num{1.2e-3}$ has been determined, we can plot some example cost spaces for various $true$ \glspl{mua} and \glspl{musp}. This is shown for \num{9} cases in Figure~\ref{fig:costSpace}. For~the \num{9} cases all combinations of the following optical properties were used: $\acrshort{mua}=$\SIlist{0.005;0.010;0.020}{\per\milli\meter} combined with $\acrshort{musp}=$\SIlist{0.5;1.0;2.0}{\per\milli\meter}. \par

Examining Figure~\ref{fig:costSpace} we notice that in general \gls{mua} will likely have more error or has a less unique solution compared to \gls{musp}. This is evident by the spreading of the low values of $\chi^2$ along the \gls{mua} direction near the local minimum and $true$ value. This result is an extension of what was seen for one set of optical properties in Figure~\ref{fig:kappas}b. We also notice that this oblique $\chi^2$ space shape is worse for small \gls{musp} (\SI{0.5}{\per\milli\meter}), which is somewhat expected since diffusion theory is not meant to be used in the low scattering regime. For~this reason, finding optimal $\kappa$ as a function of \gls{mua} and \gls{musp} may help alleviate this problem. Regardless, from~this result we should expect the fit to work less well when attempting to retrieve \gls{mua} when \gls{musp} is~low. \par

\unskip

\subsubsection{Fit Initial~Guess}
We finish our development of the fit for \gls{mua} and \gls{musp} with a demonstration of exact retrieval when the same inverse and forward models are used for \gls{TCom} without noise (Equation~(\ref{equ:TCom})). In~doing so we also investigate the effect of different initial guesses on \gls{mua} and \gls{musp} to show that convergence is not dependent on this initial guess \hl{(}The result is not dependent of initial guess given that the initial guess is of reasonable optical properties\hl{)}. For this, the~fit was implemented by using the \gls{MATLAB} function \lstinline[style=Matlab-editor]!fmincon! to minimize $\chi^2$ (Equation~(\ref{equ:chi2}); $\kappa=\num{1.2e-3}$) function. For~\lstinline[style=Matlab-editor]!fmincon!, the~algorithm \lstinline[style=Matlab-editor]!interior-point! was used and the minimum constraints on \gls{mua} and \gls{musp} set to \lstinline[style=Matlab-editor]![0,0]!, respectively, with~all other bound types~unconstrained. \par

Using this optimization setup, the~fit was run with the $\acrshort{mua}_{,true}=\SI{0.010}{\per\milli\meter}$ and the $\acrshort{musp}_{,true}=\SI{1.0}{\per\milli\meter}$ using \num{4} different initial $guess$es:
\begin{itemize}
	\item $\acrshort{mua}_{,guess}=\SI{0.005}{\per\milli\meter}$ \& $\acrshort{musp}_{,guess}=\SI{0.5}{\per\milli\meter}$. 
	\item $\acrshort{mua}_{,guess}=\SI{0.005}{\per\milli\meter}$ \& $\acrshort{musp}_{,guess}=\SI{2.0}{\per\milli\meter}$.
	\item $\acrshort{mua}_{,guess}=\SI{0.020}{\per\milli\meter}$ \& $\acrshort{musp}_{,guess}=\SI{0.5}{\per\milli\meter}$.
	\item $\acrshort{mua}_{,guess}=\SI{0.020}{\per\milli\meter}$ \& $\acrshort{musp}_{,guess}=\SI{2.0}{\per\milli\meter}$.
\end{itemize}
\vspace{3pt}

\begin{figure}[H]
	\hspace{-6pt}\includegraphics[width=11.5 cm]{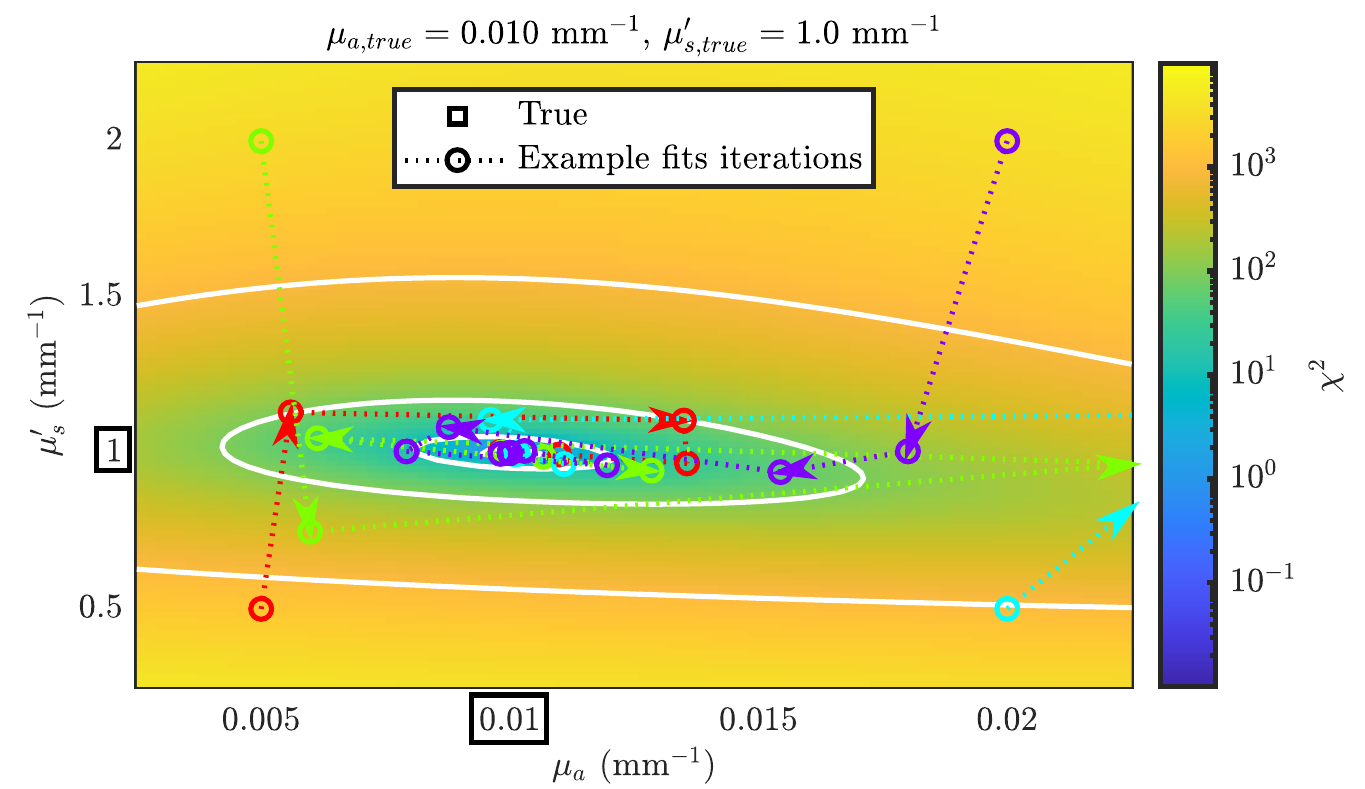}
	\caption{{Example} {trajectories} of \lstinline[style=Matlab-editor]!fmincon! minimization of $\chi^2$ (Equation~(\ref{equ:chi2}); $\kappa=\num{1.2e-3}$) to fit for the \acrshort{mua} and the \acrshort{musp}. Results from \num{4} different initial $guess$es shown:
	\textcolor[rgb]{1,0,0}{{(\textbf{Red})}} $\acrshort{mua}_{,guess}=\SI{0.005}{\per\milli\meter}$ \& $\acrshort{musp}_{,guess}=\SI{0.5}{\per\milli\meter}$. \textcolor[rgb]{0.5,1,0}{{(\textbf{Green})}} $\acrshort{mua}_{,guess}=\SI{0.005}{\per\milli\meter}$ \& $\acrshort{musp}_{,guess}=\SI{2.0}{\per\milli\meter}$. \textcolor[rgb]{0,1,1}{(\textbf{Cyan})} $\acrshort{mua}_{,guess}=\SI{0.020}{\per\milli\meter}$ \& $\acrshort{musp}_{,guess}=\SI{0.5}{\per\milli\meter}$. \textcolor[rgb]{0.5,0,1}{(\textbf{Purple})} $\acrshort{mua}_{,guess}=\SI{0.020}{\per\milli\meter}$ \& $\acrshort{musp}_{,guess}=\SI{2.0}{\per\milli\meter}$. 
	Acronyms and Symbols: Cost ($\chi^2$), scaling factor ($\kappa$), \acrfull{mua}, and~\acrfull{musp}.
	\label{fig:exampleFit}}
\end{figure}

The results from these fits and the fit trajectory (shown as dotted lines with circles) are shown in Figure~\ref{fig:exampleFit}. In~all cases the fit converged to the $true$ optical properties regardless of start point. A~second observation that can be made from Figure~\ref{fig:exampleFit} is what trajectory the fit follows during convergence. Acknowledging that this is highly dependent on algorithm choice, we still note that the fit spent most of its time traversing in the \gls{mua} direction, converging close to the correct \gls{musp} comparatively fast. This is a consequence of the shape of cost space, having a longer trough in the \gls{mua} direction than the \gls{musp}. \par

\unskip

\subsection{Confounds to Fit Retrieved Absolute Optical~Properties}\label{sec:confounds}
\unskip
\subsubsection{Propagation of Noise to Optical Property~Uncertainty}\label{sec:noiseProp}
To test how noise propagates through the recovery of \gls{mua} and \gls{musp}, when using the fit developed in Section~\ref{sec:fit}, we simulated $\sigma_{|\acrshort{TCom}|}/|\acrshort{TCom}|=\num{0.01}$ and $\sigma_{\angle\acrshort{TCom}}=\SI{1.7}{\milli\radian}=\ang{0.1}$ as mentioned in Section~\ref{sec:fitCost}. This was done by simulating measured \gls{DRTComLnI} and \gls{DRTComP} \num{101} times and each time adding Gaussian noise with the $\sigma$s stated above. For~each of the \num{101}, the~fit was run to recover some \gls{mua} and \gls{musp}. This was done for all \num{9} of the sets of $true$ \gls{mua} and \gls{musp} shown in Figure~\ref{fig:costSpace}. \par

\begin{figure}[H]
	
\begin{adjustwidth}{-\extralength}{0cm}
\centering %% If there is a figure in wide page, please release command \centering
\includegraphics{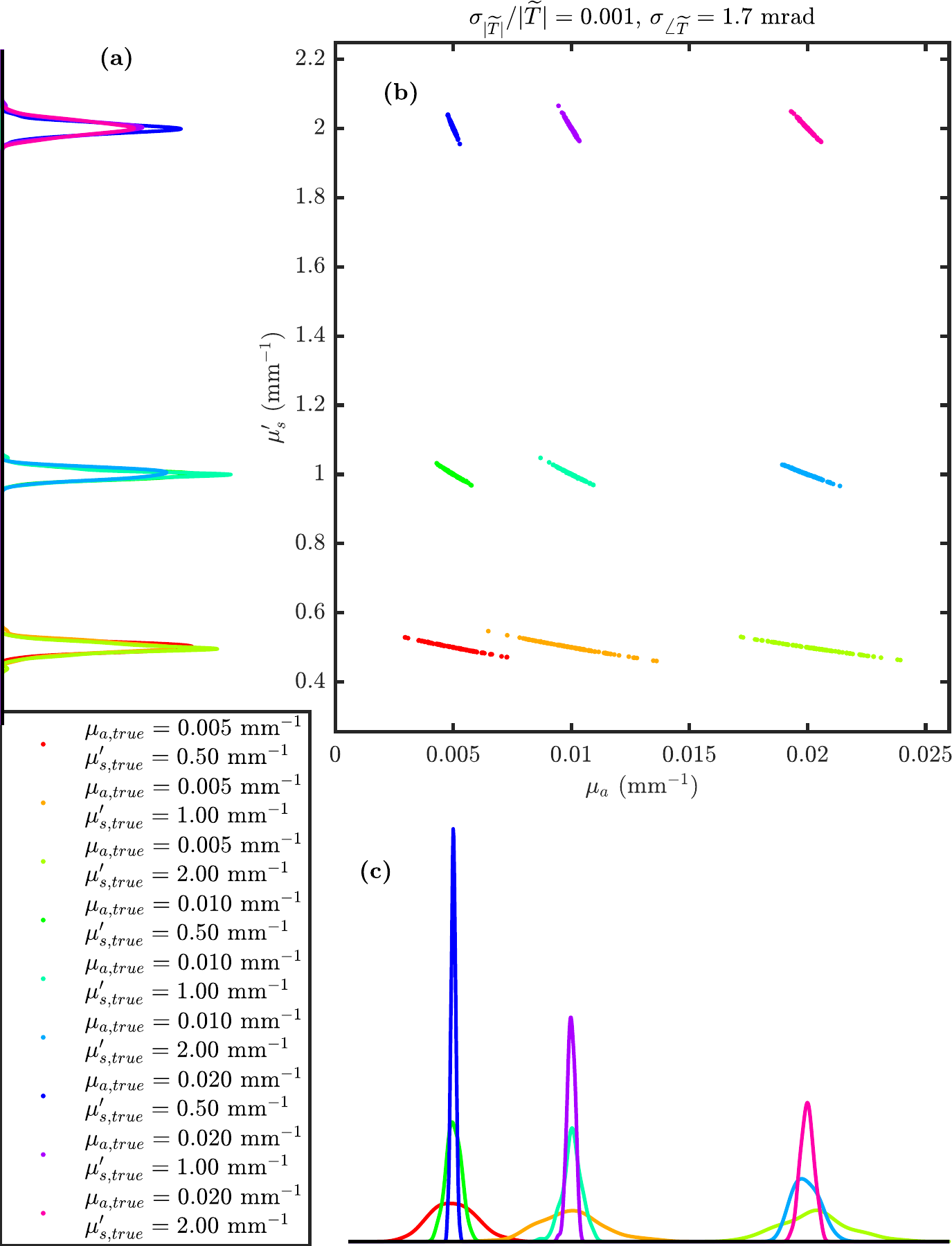}
\end{adjustwidth}
	\caption{\hl{Result from} %MDPI: the number 0.025 in the picture is not completed, please revise it.
		% GB: Fixed
 \num{101} simulations of noise ($\sigma_{|\acrshort{TCom}|}/|\acrshort{TCom}|=\num{0.01}$ and $\sigma_{\angle\acrshort{TCom}}=\SI{1.7}{\milli\radian}=\ang{0.1}$) for the \num{9} different $true$ sets of the \acrshort{mua} and the \acrshort{musp} shown in Figure~\ref{fig:costSpace}. (\textbf{a}) Marginal histograms for recovered \acrshort{musp} values of the \num{9} sets of optical properties and \num{101} noise simulations. (\textbf{b}) Scatter plot of recovered \acrshort{mua} and \acrshort{musp} for the \num{9x101} noise simulations. (\textbf{c}) Marginal histograms for recovered \acrshort{mua} values of the \num{9} sets of optical properties and \num{101} noise~simulations. 
		Acronyms and Symbols: \Acrfull{TCom}, uncertainty ($\sigma$), \acrfull{mua}, and~\acrfull{musp}.
		\label{fig:exampleErrors}}
\end{figure}

The results from this exercise are shown in Figure~\ref{fig:exampleErrors} and Table~\ref{tab:exampleErrors}, from~these three main observations can be~drawn:
\begin{enumerate}[label=\Alph*]
	\item The fractional error in \gls{mua} is always larger compared to \gls{musp} suggesting the system can more precisely recover \gls{musp}. \label{enum:muaErrBig}
	\item Errors in \gls{mua} are much larger for small \gls{musp} and slightly larger for small \gls{mua} (with small \gls{mua} and \gls{musp} together being the worst case). \label{enum:smallMuspBad}
	\item {That \gls{mua} and \gls{musp} are highly negatively correlated (as suggested by Figure~\ref{fig:kappas}{b,c}).} \label{enum:muaMuspCorr}
\end{enumerate}

Observation~\ref{enum:muaErrBig} again expounds upon what has been expected from the shape of $\chi^2$ space presented in previous sections. Furthermore, observations~\ref{enum:smallMuspBad} and \ref{enum:muaMuspCorr} are typical for such diffusion theory based~problems. \par

To draw some quantitative values for this exercise, we can closely examine Table \ref{tab:exampleErrors}. It is helpful to extract the worst case (for the type of simulations we have done), and~typical (considering typical being the case when $\acrshort{mua}_{,true}=\SI{0.010}{\per\milli\meter}$ and $\acrshort{musp}_{,true}=\SI{1.0}{\per\milli\meter}$) fractional errors in \gls{mua} and \gls{musp}. These are as~follows:
\begin{itemize}
	\item For \gls{mua}:
	\begin{itemize}
		\item Typical error of \SI{4}{\percent}.
		\item Worst case error of \SI{20}{\percent} (for low \gls{mua} and \gls{musp}).
	\end{itemize}
	\item For \gls{musp}:
	\begin{itemize}
		\item Typical error of \SI{1}{\percent}.
		\item Worst case error of \SI{3}{\percent} (for high \gls{mua} and low \gls{musp}).
	\end{itemize}
\end{itemize}

Of course these values are dependent on the simulated measurement errors of $\sigma_{|\acrshort{TCom}|}/|\acrshort{TCom}|=\num{0.01}$ and $\sigma_{\angle\acrshort{TCom}}=\SI{1.7}{\milli\radian}=\ang{0.1}$ which may be different for different~instruments. \par

\begin{table}[H]
	\centering
	\caption{Errors for \num{9} sets of $true$ optical properties and \num{101} noise simulations using $\sigma_{|\acrshort{TCom}|}/|\acrshort{TCom}|=\num{0.01}$ and $\sigma_{\angle\acrshort{TCom}}=\SI{1.7}{\milli\radian}=\ang{0.1}$.\label{tab:exampleErrors}}
	\begin{threeparttable}
	\setlength{\tabcolsep}{4.1mm}
		\begin{tabular}{S[table-format=1.3] S[table-format=1.1]  S[table-format=1.4] S[table-format=1.2]  S[table-format=1.2] S[table-format=1.2]  S[table-format=2.4]}
		\toprule
			\boldmath{$\acrshort{mua}_{,true}$} &
			 \boldmath{$\acrshort{musp}_{,true}$} &
			  \boldmath{$\sigma_{\acrshort{mua}}$} &
			\boldmath{$\sigma_{\acrshort{mua}}/\bar{\acrshort{mua}}$} &
			\boldmath{$\sigma_{\acrshort{musp}}$} &
			\boldmath{$\sigma_{\acrshort{musp}}/\bar{\acrshort{musp}}$} & \boldmath{$r_{\acrshort{mua},\acrshort{musp}}$} \\
			\textbf{(mm\textsuperscript{$-$1})} &
			\textbf{(mm\textsuperscript{$-$1})} &
			\textbf{(mm\textsuperscript{$-$1})} &
			{} &
			\textbf{(mm\textsuperscript{$-$1})} &
			{} &
			{} \\
			
			\midrule
			0.005 & 0.5 & 0.0008  & 0.2  & 0.01 & 0.02  & -0.9986\\
			0.005 & 1.0 & 0.0003  & 0.06 & 0.01 & 0.01  & -0.9970\\
			0.005 & 2.0 & 0.0001  & 0.02 & 0.02 & 0.008 & -0.9931\\
			0.010 & 0.5 & 0.001   & 0.1  & 0.01 & 0.03  & -0.9986\\
			0.010 & 1.0 & 0.0004  & 0.04 & 0.01 & 0.01  & -0.9983\\
			0.010 & 2.0 & 0.0002  & 0.02 & 0.02 & 0.009 & -0.9971\\
			0.020 & 0.5 & 0.001   & 0.07 & 0.02 & 0.03  & -0.9990\\
			0.020 & 1.0 & 0.0007  & 0.04 & 0.02 & 0.02  & -0.9988\\
			0.020 & 2.0 & 0.0003  & 0.01 & 0.02 & 0.009 & -0.9981\\
			\bottomrule
		\end{tabular}
		\noindent{\footnotesize{Symbols: \Acrfull{mua}, \acrfull{musp}, \acrfull{TCom}, uncertainty ($\sigma$), and~correlation coefficient ($r$).}}
%	\begin{tablenotes}
%		\item Symbols: \Acrfull{mua}, \acrfull{musp}, \acrfull{TCom}, uncertainty ($\sigma$), and~correlation coefficient ($r$).
%	\end{tablenotes}
	\end{threeparttable}
\end{table}
\unskip

\subsubsection{Assumption of  Index of Refraction}\label{sec:ass_n}
\begin{figure}[H]
	\includegraphics{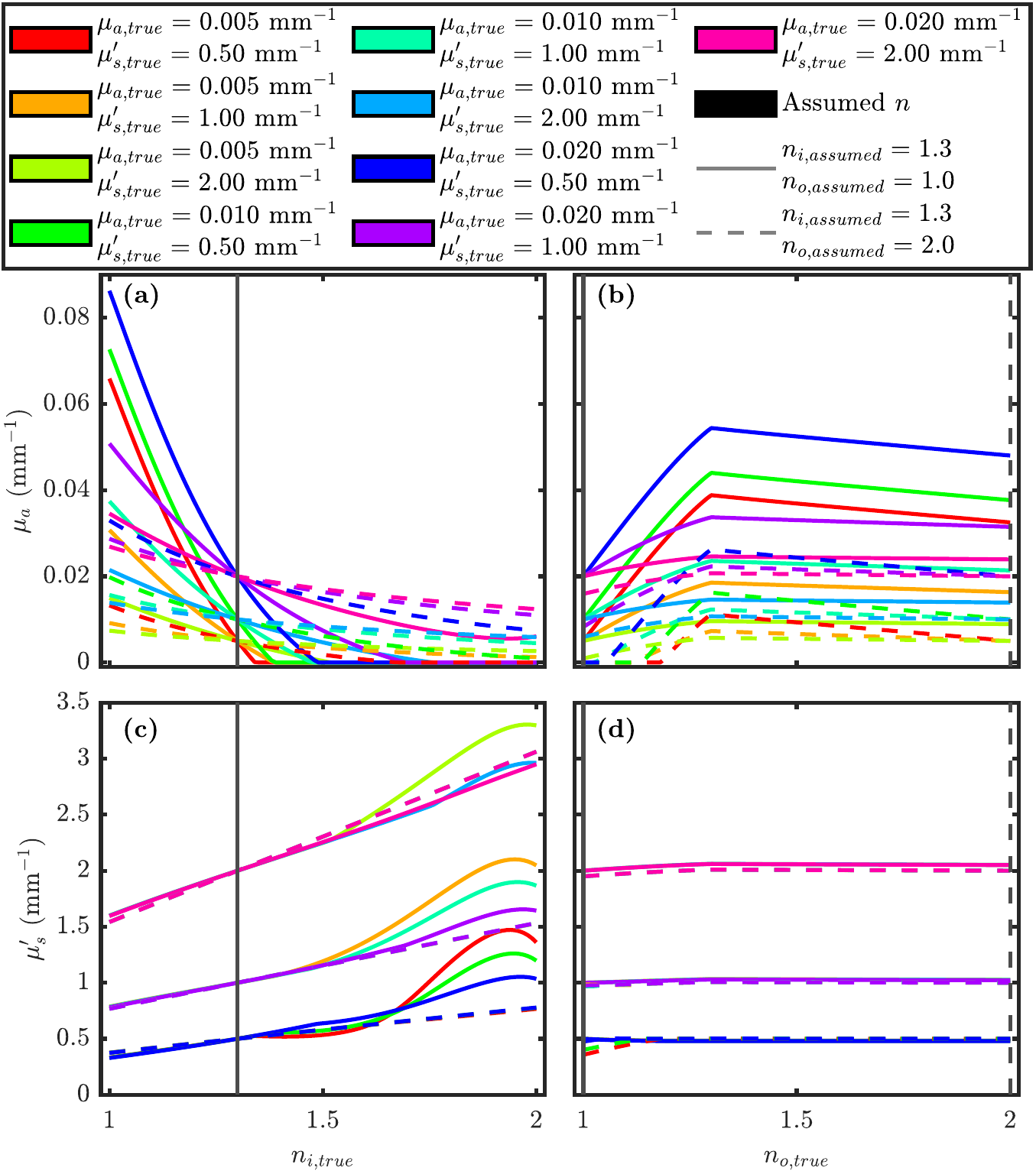}
	\caption{Effect of recovered \acrshort{mua} and \acrshort{musp} on the $true$ \acrshort{n} inside ($\acrshort{n}_{i}$) and outside ($\acrshort{n}_{o}$) when fixed values of \acrshortpl{n} are $assumed$. Shown for the \num{9} sets of optical properties used in Figure~\ref{fig:costSpace}. (\textbf{Solid Lines}) $\acrshort{n}_i=\num{1.3}$ and $\acrshort{n}_o=\num{1.0}$ $assumed$ in fit. (\textbf{Dashed Lines}) $\acrshort{n}_i=\num{1.3}$ and $\acrshort{n}_o=\num{2.0}$ $assumed$ in fit. (\textbf{a}) Recovered \acrshort{mua} while varying $\acrshort{n}_{i,true}$ and fixing $\acrshort{n}_{o,true}$ to the $assumed$ value. (\textbf{b}) Recovered \acrshort{mua} while varying $\acrshort{n}_{o,true}$ and fixing $\acrshort{n}_{i,true}$ to the $assumed$ value.  (\textbf{c}) Recovered \acrshort{musp} while varying $\acrshort{n}_{i,true}$ and fixing $\acrshort{n}_{o,true}$ to the $assumed$ value.  (\textbf{d}) Recovered \acrshort{musp} while varying $\acrshort{n}_{o,true}$ and fixing $\acrshort{n}_{i,true}$ to the $assumed$ value. 
		Acronyms and Symbols: \Acrfull{mua}, \acrfull{musp}, and~\acrfull{n}.
		\label{fig:wrongIndex}}
\end{figure}

Finally, we examine how the assumption of \gls{n} (and by extension the model boundary conditions) affects the recovered \gls{mua} and \gls{musp}. We have done this by running the fit \hl{$assum$ing} %MDPI: Is the italic necessary? please check it through the whole maintext.
% GB: Yes we would like to keep it.
 sets of $\acrshort{n}_i$ and $\acrshort{n}_o$, but~generating forward data with different \hl{$true$} \glspl{n} in the range \numrange{1}{2} (we do not co-vary $\acrshort{n}_i$ and $\acrshort{n}_o$ for simplicity). \num{2} sets of \glspl{n} \hl{$assumed$} in the fit were~invested:
\begin{itemize}
	\item $\acrshort{n}_{i,assumed}=\num{1.3}$ \& $\acrshort{n}_{o,assumed}=\num{1.0}$ (Figure~\ref{fig:wrongIndex} solid lines).
	\item $\acrshort{n}_{i,assumed}=\num{1.3}$ \& $\acrshort{n}_{o,assumed}=\num{2.0}$ (Figure~\ref{fig:wrongIndex} dashed lines).
\end{itemize}
\vspace{3pt}

This exercise was done for all \num{9} sets of \gls{mua} and \gls{musp} shown in Figure~\ref{fig:costSpace}. \par

Figure~\ref{fig:wrongIndex} shows these recovered \gls{mua} and \gls{musp} for the \num{2} $assumed$ cases while varying $\acrshort{n}_{i,true}$ and $\acrshort{n}_{o,true}$. First, we note that $\acrshort{n}_i$ has a larger effect on the recovered \gls{mua} and \gls{musp} compared to $\acrshort{n}_o$, with~\gls{mua} having a negative, and~\gls{musp} a positive, correlation with $\acrshort{n}_{i,true}$ (Figure~\ref{fig:wrongIndex}{a,c}). Furthermore, \gls{mua} is much more strongly affected by $\acrshort{n}_{i,true}$ than \gls{musp}, with~recovered values being up to about \num{7} times greater than the $true$ value when there is a low $\acrshort{n}_{i,true}$ value. For~high $\acrshort{n}_{i,true}$ the recovered \gls{mua} often approaches \SI{0}{\per\milli\meter} (hitting the \lstinline[style=Matlab-editor]!fmincon! constraint). All of this suggests that the method's ability to accurately recover \gls{musp} and particularly \gls{mua} is dependent on knowledge of $\acrshort{n}_i$. \par

Now, focusing on Figure~\ref{fig:wrongIndex}{b,d}, we see the effect of $\acrshort{n}_o$. In~this case \gls{musp} is almost not affected at all by $\acrshort{n}_{o,true}$ and \gls{mua} is much more significantly affected when $\acrshort{n}_{o,true}<\acrshort{n}_{i,true}$ or $\acrshort{n}_{r,true}>1$. In~this case the correlation between \gls{mua} and $\acrshort{n}_o$ is positive (opposite to that for $\acrshort{n}_i$), suggesting a connection to the dependence on $\acrshort{n}_r$. These results further re-enforce the idea that recovered \gls{mua} would be highly effected by the true \glspl{n}, thus control or knowledge of the \glspl{n} for this method is~critical. \par

Lastly, by~comparing the two $\acrshort{n}_{i,assumed}$ and $\acrshort{n}_{o,assumed}$ sets (solid versus dashed lines) we notice that the recovered \gls{mua} and \gls{musp} vary less for the dashed lines (Figure~\ref{fig:wrongIndex}). The~dashed line is the case where $\acrshort{n}_{i,assumed}=\num{1.3}$ and $\acrshort{n}_{o,assumed}=\num{2.0}$ \hl{(}In Figure~\ref{fig:wrongIndex} the $assumed$ \acrfull{n}, is equal to the $true$ when the other \acrshort{n} is varied, for~example in Figure~\ref{fig:wrongIndex}a the true \acrshort{n} outside ($\acrshort{n}_{o,true}$) is \num{1} for the solid line and \num{2} for the dashed line\hl{)}. This tells us that the incorrect recovery of \gls{mua} and \gls{musp} can be partially alleviated when $\acrshort{n}_o$ is large, even if $\acrshort{n}_i$ in unknown. Since when this method is implemented it would be more practical to control $\acrshort{n}_o$ than $\acrshort{n}_i$, it would be advantageous to design a cuvette with high \gls{n} to take advantage of this reduction of the effect of the assumption of $\acrshort{n}_i$ seen by comparing dashed to solid lines in Figure~\ref{fig:wrongIndex}a. \par

%%%%%%%%%%%%%%%%%%%%%%%%%%%%%%%%%%%%%%%%%%
\section{Discussion}\label{sec:dis}
The method presented appears to be feasible in measuring absolute \gls{mua} and \gls{musp} in a standard cuvette (\SI{45x10x10}{\milli\meter}). This is significant given that typical/traditional measurements of \gls{mua} and \gls{musp} with diffuse optical methods require large sample volumes (on the order of \si{liters}) and careful instrumental calibration. In~this case, small samples volumes may be used (on the order of \SIrange{1}{10}{\milli\liter}) without the need for calibration of optode coupling (as described in Section~\ref{sec:selfCal}). \par

To summarize, we started the development of this measurement method by choosing which data we intended to collect from the cuvette, namely \gls{DRTComLnI} and \gls{DRTComP}, and~determining how these data vary in respect to the desired recovered properties, namely \gls{mua} and \gls{musp} (Figure~\ref{fig:diffGrd}). This leads to the development of a fit for \gls{mua} and \gls{musp} and a careful examination of $\chi^2$ space (Section~\ref{sec:fit}). From~this examination, one major result was discovered, this being that \gls{mua} is less determined (has a broad local minimum area in $\chi^2$ space) compared to \gls{musp}. This is the first potential limitation of this method since often \gls{mua} is in-fact the targeted property of interest while \gls{musp} may be considered a confound. Despite this it appears that this weakness mainly occurs when \gls{musp} is small (\SI{<1}{\per\milli\meter}; Figure~\ref{fig:costSpace}) telling us that the method has its main strength when the sample is highly scattering. {Given that most commercial spectrometers designed for cuvette measurement require a non-scattering sample.} \par

We also simulated two types of confounds that may lead to incorrect recovered \gls{mua} and \gls{musp}. First, we investigated how instrumental noise would propagate through the measurement to the recovered \gls{mua} and \gls{musp} (Section~\ref{sec:noiseProp}). Here we confirmed what was expected when the $\chi^2$ space was examined, specifically that \gls{mua} has a higher relative error compared to \gls{musp} (Figure~\ref{fig:exampleErrors} and Table~\ref{tab:exampleErrors}). However, this error becomes comparable when \gls{musp} is high. For~example, with~$\sigma_{|\acrshort{TCom}|}/|\acrshort{TCom}|=\num{0.01}$ and $\sigma_{\angle\acrshort{TCom}}=\SI{1.7}{\milli\radian}=\ang{0.1}$:
\begin{itemize}
	\item If $\acrshort{mua}=\SI{0.005}{\per\milli\meter}$ \& $\acrshort{musp}=\SI{0.5}{\per\milli\meter}$ then \gls{mua} has an error of \SI{20}{\percent} and \gls{musp} of \SI{2}{\percent}.
	\item If $\acrshort{mua}=\SI{0.020}{\per\milli\meter}$ \& $\acrshort{musp}=\SI{2.0}{\per\milli\meter}$ then \gls{mua} has an error of \SI{1}{\percent} and \gls{musp} of \SI{0.9}{\percent}.
\end{itemize}

Therefore, we see that this method really shines when the sample is very diffuse, which is another way of saying highly~scattering. \par

Investigation of the effect of incorrectly assumed boundary conditions on the fit results was also done (Section~\ref{sec:ass_n}).
\q{R2C5}{{Many different boundary conditions have been extensively studied and modeled in the past~\cite{pattersonTimeResolvedReflectance1989, aronsonBoundaryConditionsDiffusion1995, popescuEvidenceScatteringAnisotropy2000, ripollBoundaryConditionsLight2000, el-wakilRadiativeTransferSpherical2001}, however just because conditions can be modeled does not mean that there is knowledge of them in practice which can be corrected for.}}
Given that a diffusion theory model was used, we varied boundary conditions in terms of $\acrshort{n}_i$ and $\acrshort{n}_o$. Again, we found that \gls{mua} is more likely to be incorrectly recovered compared to \gls{musp}. However, further we found that $\acrshort{n}_i$ had the largest effect on the recovered \gls{mua} and \gls{musp} (Figure~\ref{fig:wrongIndex}). This in principle is a short-coming of the method since one could argue that if the method were implemented, $\acrshort{n}_o$ could be controlled through instrument design but $\acrshort{n}_i$ would be unknown. However, Figure~\ref{fig:wrongIndex} shows that the effect of $\acrshort{n}_i$ is suppressed when $\acrshort{n}_o$ is large, suggesting a relationship to $\acrshort{n}_r$. Therefore, we expect that an instrumental design for this method would include a cuvette designed for high $\acrshort{n}_o$. 
\q{R4C3}{{Additionally, the~current model considers a cuvette closed on all sides so that $\acrshort{n}_{o}$ is the same on all six faces. This is unrealistic for a typical cuvette which would have one side open to the air. When this method is implemented in practice, either a lid with the same material as the cuvette would need to be incorporated or the air boundary on one side considered. If~a top air boundary is considered the \glspl{SRTCom} would not be the same in theory, but~the \gls{SC}/\gls{DS} coupling cancellation would still apply. The~model would need to be more complex since \gls{DRTCom} would no longer equal the theoretical \gls{SRTCom} but instead the average of the two since theoretical \glspl{SRTCom} would not be equal, but~given the correct model an inversion is still expected to work.}}
Further, we also note that examining the expression for \gls{mueffCom} (Equation~(\ref{equ:mueffCom})) ones sees that \gls{mua} and \gls{musp} are coupled to $\acrshort{n}_i$. This means that any diffuse measurement using such theory would actually measure $\acrshort{musp}\acrshort{n}_i$ and $\acrshort{mua}/\acrshort{n}_i$. Therefore, cross-talk with $\acrshort{n}_i$ is a necessary consequence of the theory and can only really be suppressed, not removed. This is seen by examining the data-sheet for the \gls{SS150H} which utilizes the integrating sphere measurement method~\cite{foschumPreciseDeterminationOptical2020, bergmannPreciseDeterminationOptical2020}. The~\gls{SS150H} states that with a \gls{n} uncertainty of \num{0.06} one should expect a \gls{mua} uncertainty of \SI{12}{\percent} and a \gls{musp} of \SI{7}{\percent} \cite{optikSphereSpectro150H}. Therefore, the~method presented here is, at~least in theory, comparable to existing~instruments. \par

Finally, we revisit the idea of calibration. In~Section~\ref{sec:selfCal}, we showed that this method takes the advantages of \gls{SC} \cite{hueberNewOpticalProbe1999}/\gls{DS} \cite{sassaroliDualslopeMethodEnhanced2019, blaneyPhaseDualslopesFrequencydomain2020} meaning that the measurements of \gls{DRTComLnI} and \gls{DRTComP} are insensitive to instrumental coupling. However, due to the fact that this method utilizes a small geometry and is highly affected by boundary conditions, other instrumental calibration may be required. First, since this diffusion theory solution (Equation~(\ref{equ:TCom})) is for such a small geometry, calibration of the inverse model maybe necessary if factors exist which are not modeled by \gls{CCom} (Section~\ref{sec:selfCal}). Three options are available for creating an inverse~model: 
\begin{enumerate}[label=\Roman*]
	\item Diffusion theory based cost minimization (shown here). \label{enum:DTfit}
	\item Look-up table with Monte-Carlo generated data. \label{enum:MCgenLU}
	\item Look-up table with instrumental measurements of known samples. \label{enum:expGenLU}
\end{enumerate}

Option~\ref{enum:DTfit} is the most elegant which is why it was chosen here, but~option~\ref{enum:expGenLU} (being the most brute-force and likely infeasible in practice due to the extensive calibration phantom preparation and measurement needed for each unique instrument) would almost definitely work, and~allows for correction of systematic confounds, provided that the measurements are repeatable. The~auto-calibration in Section~\ref{sec:selfCal} is expected to significantly help with this repeatability, but~the biggest secondary factor is repeatable boundary conditions. The~investigation here showed promise to alleviate the boundary conditions issue by using high $\acrshort{n}_o$, but~future work will investigate the repeatability of measurements of cuvettes with various boundary conditions experimentally.
\q{R3}{{Of course, this future work would involve experimental implementation of the measurement. For~this we plan to utilize the \gls{ISSv2} which utilizes fiber bundles that will be coupled to the sides of a standard cuvette. Regardless of the coupling method, we expect the \gls{SC}/\gls{DS} method to compensate for optode and coupling losses.}}
\q{R4C4}{{However, these optodes will not act as true pencil beams or point detectors as is modeled here. This is another condition which could cause errors in the measurement and would need to be modeled or calibrated for. Area detectors may be modeled with diffusion theory by integrating over the area of the detector while various types of sources could be modeled using Monte-Carlo. If~such realistic forward models are still not enough to account for the practicals of the optodes themselves then option~\ref{enum:expGenLU} may be needed. However, we emphasize that we expect that these considerations will be partially alleviated by the \gls{SC}/\gls{DS} method and its measurement symmetry.}} \par

%%%%%%%%%%%%%%%%%%%%%%%%%%%%%%%%%%%%%%%%%%
\section{Conclusions}
The purpose of this article is to present and determine the feasibility as well as strengths and weaknesses of a method to measure diffuse absolute optical properties in a standard cuvette. The~strengths of this method lie in the way it is posited, a~way to measure absolute diffuse optical properties in small samples for which no commercial instruments which utilize frequency-domain type measurements exist to our knowledge. Our intention is to expand this method to spectral measurements of absorption to recover chemical concentrations of a diffuse sample~\cite{blaneyBroadbandAbsorptionSpectroscopy2021}. Two main limitations were found in this method: first, higher error in absorption properties compared to scattering; second, high dependence on the knowledge of the index of refraction of the sample. However, the~investigation lead to possible methods to address or alleviate these limitations. {That being, the~measurement of strongly scattering samples to address the first limitation, and~the use of a cuvette with high index of refraction to address the second.} Future work will move beyond theoretical development of the method to experimental implementation, and~an investigation of boundary conditions and repeatability which can really only be done in experimental~practice. \par

%%%%%%%%%%%%%%%%%%%%%%%%%%%%%%%%%%%%%%%%%%
\vspace{6pt} 
%%%%%%%%%%%%%%%%%%%%%%%%%%%%%%%%%%%%%%%%%%
\authorcontributions{Conceptualization, G.B., A.S. and~S.F.; methodology, G.B. and A.S.; software, G.B. and A.S.; validation, G.B. and A.S.; formal analysis, G.B.; investigation, G.B.; resources, A.S. and S.F.; data curation, G.B.; writing---original draft preparation, G.B.; writing---review and editing, G.B., A.S. and~S.F.; visualization, G.B.; supervision, A.S. and S.F.; project administration, S.F.; funding acquisition, S.F. All authors have read and agreed to the published version of the~manuscript.}

\funding{\hl{This research} %MDPI: Please check carefully the funding data and funding numbers. GB: Done
 was funded by \acrfull{NIH} grant numbers R01-NS095334 and~R01-EB029414.}
\institutionalreview{\hl{Not applicable}}%MDPI: In this section, you should add the Institutional Review Board Statement and approval number, if relevant to your study. You might choose to exclude this statement if the study did not require ethical approval. Please note that the Editorial Office might ask you for further information. Please add “The study was conducted in accordance with the Declaration of Helsinki, and approved by the Institutional Review Board (or Ethics Committee) of NAME OF INSTITUTE (protocol code XXX and date of approval).” for studies involving humans. OR “The animal study protocol was approved by the Institutional Review Board (or Ethics Committee) of NAME OF INSTITUTE (protocol code XXX and date of approval).” for studies involving animals. OR “Ethical review and approval were waived for this study due to REASON (please provide a detailed justification).” OR “Not applicable” for studies not involving humans or animals.
% GB: Added not applicable

\informedconsent{\hl{Not applicable}}%MDPI:Any research article describing a study involving humans should contain this statement. Please add ``Informed consent was obtained from all subjects involved in the study.'' OR ``Patient consent was waived due to REASON (please provide a detailed justification).'' OR ``Not applicable'' for studies not involving humans. You might also choose to exclude this statement if the study did not involve humans.
% GB: Added not applicable

%Written informed consent for publication must be obtained from participating patients who can be identified (including by the patients themselves). Please state ``Written informed consent has been obtained from the patient(s) to publish this paper'' if applicable.
\dataavailability{\url{https://github.com/DOIT-Lab/DOIT-Public/tree/master/OpticalPropertiesInCuvette} (\hl{25 October 2022}). %MDPI: Please add the access date (format: Date Month Year), e.g., accessed on 1 January 2020. GB: Done
}

\conflictsofinterest{The authors declare a current patent application regarding the method presented in this~article.} 

%%%%%%%%%%%%%%%%%%%%%%%%%%%%%%%%%%%%%%%%%%
\glsresetall
\renewcommand{\glossarysection}[2][]{}
\abbreviations{Symbols}{
	\hl{The following symbols are used in this manuscript}: %MDPI: please confirm the format of the symbolslist below, is the page number and link of it necessary? If you want to keep this format, please write the tex code below, not only ``\printglossary[type=symbolslist]'', otherwise, we cannot continue the next step, thanks.
	% GB: If you can not use \printglossary simply excude the list of symbols. If you can keep it without the page link that is better. Thus, in case it is the numbers that are the problem I have removed them with ``nonumberlist''.

	\noindent 
%	\printglossary[type=\acronymtype]
	\printglossary[type=symbolslist,nonumberlist]
	
}

%%%%%%%%%%%%%%%%%%%%%%%%%%%%%%%%%%%%%%%%%%
\begin{adjustwidth}{-\extralength}{0cm}
\reftitle{References}

\end{adjustwidth}

\begin{thebibliography}{999}

\bibitem[Bigio and Fantini(2016)]{bigioQuantitativeBiomedicalOptics2016}
Bigio, I.J.; Fantini, S.
\newblock {\em Quantitative {{Biomedical Optics}}: {{Theory}}, {{Methods}}, and
  {{Applications}}}; {Cambridge University Press: Cambridge, UK},  2016.

\bibitem[Quaresima and
  Ferrari(2019)]{quaresimaFunctionalNearInfraredSpectroscopy2019}
Quaresima, V.; Ferrari, M.
\newblock Functional {{Near-Infrared Spectroscopy}} ({{fNIRS}}) for {{Assessing
  Cerebral Cortex Function During Human Behavior}} in {{Natural}}/{{Social
  Situations}}: {{A Concise Review}}.
\newblock {\em Organ. Res. Methods} {\bf 2019}, {\em 22},~46--68.
\newblock {{https://doi.org/10.1177/1094428116658959}}.

\bibitem[Rahman \em{et~al.}(2020)Rahman, Siddik, Ghosh, Khanam, and
  Ahmad]{rahmanNarrativeReviewClinical2020}
Rahman, M.A.; Siddik, A.B.; Ghosh, T.K.; Khanam, F.; Ahmad, M.
\newblock A {{Narrative Review}} on {{Clinical Applications}} of {{fNIRS}}.
\newblock {\em J. Digit. Imaging} {\bf 2020}, {\em 33},~1167--1184.
\newblock {{https://doi.org/10.1007/s10278-020-00387-1}}.

\bibitem[Ozaki \em{et~al.}(2006)Ozaki, McClure, and
  Christy]{ozakiNearInfraredSpectroscopyFood2006}
Ozaki, Y.; McClure, W.F.; Christy, A.A. (Eds.)
\newblock {\em Near-{{Infrared Spectroscopy}} in {{Food Science}} and
  {{Technology}}}; {John Wiley \& Sons}:  \hl{Hoboken, NJ, USA}, %newly added information, please confirm GB: Okay
  2006.

\bibitem[Johnson(2020)]{johnsonOverviewNearinfraredSpectroscopy2020}
Johnson, J.B.
\newblock An Overview of Near-Infrared Spectroscopy ({{NIRS}}) for the
  Detection of Insect Pests in Stored Grains.
\newblock {\em J. Stored Prod. Res.} {\bf 2020}, {\em
  86},~101558.
\newblock {{https://doi.org/10.1016/j.jspr.2019.101558}}.

\bibitem[Kademi \em{et~al.}(2019)Kademi, Ulusoy, and
  Hecer]{kademiApplicationsMiniaturizedPortable2019}
Kademi, H.I.; Ulusoy, B.H.; Hecer, C.
\newblock Applications of Miniaturized and Portable near Infrared Spectroscopy
  ({{NIRS}}) for Inspection and Control of Meat and Meat Products.
\newblock {\em Food Rev. Int.} {\bf 2019}, {\em 35},~201--220.
\newblock {{https://doi.org/10.1080/87559129.2018.1514624}}.

\bibitem[Razuc \em{et~al.}(2019)Razuc, Grafia, Gallo, {Ram{\'i}rez-Rigo}, and
  Roma{\~n}ach]{razucNearinfraredSpectroscopicApplications2019}
Razuc, M.; Grafia, A.; Gallo, L.; {Ram{\'i}rez-Rigo}, M.V.; Roma{\~n}ach, R.J.
\newblock Near-Infrared Spectroscopic Applications in Pharmaceutical Particle
  Technology.
\newblock {\em Drug Dev. Ind. Pharm.} {\bf 2019}, {\em
  45},~1565--1589.
\newblock {{https://doi.org/10.1080/03639045.2019.1641510}}.

\bibitem[Stranzinger \em{et~al.}(2021)Stranzinger, Markl, Khinast, and
  Paudel]{stranzingerReviewSensingTechnologies2021}
Stranzinger, S.; Markl, D.; Khinast, J.G.; Paudel, A.
\newblock Review of Sensing Technologies for Measuring Powder Density
  Variations during Pharmaceutical Solid Dosage Form Manufacturing.
\newblock {\em TrAC Trends Anal. Chem.} {\bf 2021}, {\em
  135},~116147.
\newblock {{https://doi.org/10.1016/j.trac.2020.116147}}.

\bibitem[Chen \em{et~al.}(2017)Chen, Gu, Zhang, and
  Zhang]{chenAuthenticationInferenceSeal2017}
Chen, Z.; Gu, A.; Zhang, X.; Zhang, Z.
\newblock Authentication and Inference of Seal Stamps on {{Chinese}}
  Traditional Painting by Using Multivariate Classification and Near-Infrared
  Spectroscopy.
\newblock {\em Chemom. Intell. Lab. Syst.} {\bf 2017},
  {\em 171},~226--233.
\newblock {{https://doi.org/10.1016/j.chemolab.2017.10.017}}.

\bibitem[Trant \em{et~al.}(2020)Trant, Kristiansen, and
  Sindb{\ae}k]{trantVisibleNearinfraredSpectroscopy2020}
Trant, P.L.K.; Kristiansen, S.M.; Sindb{\ae}k, S.M.
\newblock Visible Near-Infrared Spectroscopy as an Aid for Archaeological
  Interpretation.
\newblock {\em Archaeol. Anthropol. Sci.} {\bf 2020}, {\em
  12},~280.
\newblock {{https://doi.org/10.1007/s12520-020-01239-3}}.

\bibitem[Tsuchikawa and Kobori(2015)]{tsuchikawaReviewRecentApplication2015}
Tsuchikawa, S.; Kobori, H.
\newblock A Review of Recent Application of near Infrared Spectroscopy to Wood
  Science and Technology.
\newblock {\em J. Wood Sci.} {\bf 2015}, {\em 61},~213--220.
\newblock {{https://doi.org/10.1007/s10086-015-1467-x}}.

\bibitem[Fantini and
  Sassaroli(2020)]{fantiniFrequencyDomainTechniquesCerebral2020}
Fantini, S.; Sassaroli, A.
\newblock Frequency-{{Domain Techniques}} for {{Cerebral}} and {{Functional
  Near-Infrared Spectroscopy}}.
\newblock {\em Front. Neurosci.} {\bf 2020}, {\em 14},~300.
\newblock {{https://doi.org/10.3389/fnins.2020.00300}}.

\bibitem[Torricelli \em{et~al.}(2014)Torricelli, Contini, Pifferi, Caffini, Re,
  Zucchelli, and Spinelli]{torricelliTimeDomainFunctional2014}
Torricelli, A.; Contini, D.; Pifferi, A.; Caffini, M.; Re, R.; Zucchelli, L.;
  Spinelli, L.
\newblock Time Domain Functional {{NIRS}} Imaging for Human Brain Mapping.
\newblock {\em NeuroImage} {\bf 2014}, {\em 85},~28--50.
\newblock {{https://doi.org/10.1016/j.neuroimage.2013.05.106}}.

\bibitem[Foschum \em{et~al.}(2020)Foschum, Bergmann, and
  Kienle]{foschumPreciseDeterminationOptical2020}
Foschum, F.; Bergmann, F.; Kienle, A.
\newblock Precise Determination of the Optical Properties of Turbid Media Using
  an Optimized Integrating Sphere and Advanced {{Monte Carlo}} Simulations.
  {{Part}} 1: Theory.
\newblock {\em Appl. Opt.} {\bf 2020}, {\em 59},~3203--3215.
\newblock {{https://doi.org/10.1364/AO.386011}}.

\bibitem[Bergmann \em{et~al.}(2020)Bergmann, Foschum, Zuber, and
  Kienle]{bergmannPreciseDeterminationOptical2020}
Bergmann, F.; Foschum, F.; Zuber, R.; Kienle, A.
\newblock Precise Determination of the Optical Properties of Turbid Media Using
  an Optimized Integrating Sphere and Advanced {{Monte Carlo}} Simulations.
  {{Part}} 2: Experiments.
\newblock {\em Appl. Opt.} {\bf 2020}, {\em 59},~3216--3226.
\newblock {{https://doi.org/10.1364/AO.385939}}.

\bibitem[Contini \em{et~al.}(1997)Contini, Martelli, and
  Zaccanti]{continiPhotonMigrationTurbid1997}
Contini, D.; Martelli, F.; Zaccanti, G.
\newblock Photon Migration through a Turbid Slab Described by a Model Based on
  Diffusion Approximation. {{I}}. {{Theory}}.
\newblock {\em Appl. Opt.} {\bf 1997}, {\em 36},~4587--4599.
\newblock {{https://doi.org/10.1364/AO.36.004587}}.

\bibitem[Fang and Boas(2009)]{fangMonteCarloSimulation2009}
Fang, Q.; Boas, D.A.
\newblock Monte {{Carlo Simulation}} of {{Photon Migration}} in {{3D Turbid
  Media Accelerated}} by {{Graphics Processing Units}}.
\newblock {\em Opt. Express} {\bf 2009}, {\em 17},~20178--20190.
\newblock {{https://doi.org/10.1364/OE.17.020178}}.

\bibitem[Hielscher \em{et~al.}(2004)Hielscher, Klose, Scheel, {Moa-Anderson},
  Backhaus, Netz, and Beuthan]{hielscherSagittalLaserOptical2004}
Hielscher, A.H.; Klose, A.D.; Scheel, A.K.; {Moa-Anderson}, B.; Backhaus, M.;
  Netz, U.; Beuthan, J.
\newblock Sagittal Laser Optical Tomography for Imaging of Rheumatoid Finger
  Joints.
\newblock {\em Phys. Med. Biol.} {\bf 2004}, {\em
  49},~1147--1163.
\newblock {{https://doi.org/10.1088/0031-9155/49/7/005}}.

\bibitem[Klose \em{et~al.}(2005)Klose, Ntziachristos, and
  Hielscher]{kloseInverseSourceProblem2005}
Klose, A.D.; Ntziachristos, V.; Hielscher, A.H.
\newblock The Inverse Source Problem Based on the Radiative Transfer Equation
  in Optical Molecular Imaging.
\newblock {\em J. Comput. Phys.} {\bf 2005}, {\em
  202},~323--345.
\newblock {{https://doi.org/10.1016/j.jcp.2004.07.008}}.

\bibitem[Klose and Larsen(2006)]{kloseLightTransportBiological2006}
Klose, A.D.; Larsen, E.W.
\newblock Light Transport in Biological Tissue Based on the Simplified
  Spherical Harmonics Equations.
\newblock {\em J. Comput. Phys.} {\bf 2006}, {\em
  220},~441--470.
\newblock {{https://doi.org/10.1016/j.jcp.2006.07.007}}.

\bibitem[Ho \em{et~al.}(2012)Ho, Chin, Dong, and
  Lee]{hoMultiHarmonicHomodyneApproach2012}
Ho, J.H.; Chin, H.L.; Dong, J.; Lee, K.
\newblock Multi-{{Harmonic}} Homodyne Approach for Optical Property Measurement
  of Turbid Medium in Transmission Geometry.
\newblock {\em Opt. Commun.} {\bf 2012}, {\em 285},~2007--2011.
\newblock {{https://doi.org/10.1016/j.optcom.2011.12.018}}.

\bibitem[Taniguchi \em{et~al.}(2007)Taniguchi, Murata, and
  Okamura]{taniguchiLightDiffusionModel2007}
Taniguchi, J.; Murata, H.; Okamura, Y.
\newblock Light Diffusion Model for Determination of Optical Properties of
  Rectangular Parallelepiped Highly Scattering Media.
\newblock {\em Appl. Opt.} {\bf 2007}, {\em 46},~2649--2655.
\newblock {{https://doi.org/10.1364/AO.46.002649}}.

\bibitem[Kienle(2005)]{kienleLightDiffusionTurbidParallelepiped2005}
Kienle, A.
\newblock Light diffusion through a turbid parallelepiped.
\newblock {\em J. Opt. Soc. Am. A} {\bf 2005}, {\em 22},~1883--1888.
\newblock {{https://doi.org/10.1364/JOSAA. 22.001883}}.

\bibitem[Hueber \em{et~al.}(1999)Hueber, Fantini, Cerussi, and
  Barbieri]{hueberNewOpticalProbe1999}
Hueber, D.M.; Fantini, S.; Cerussi, A.E.; Barbieri, B.B.
\newblock New Optical Probe Designs for Absolute (Self-Calibrating) {{NIR}}
  Tissue Hemoglobin Measurements.
\newblock In \emph{Optical Tomography and Spectroscopy of
  Tissue III}: San Jose, CA, USA; {International Society for Optics and Photonics}:  Bellingham, WA, USA, %MDPI: please add the location. GB: Done
   1999; Volume  3597, pp. 618--631.
\newblock {{https://doi.org/10.1117/12.356784}}.

\bibitem[Bevilacqua \em{et~al.}(2000)Bevilacqua, Berger, Cerussi, Jakubowski,
  and Tromberg]{bevilacquaBroadbandAbsorptionSpectroscopy2000}
Bevilacqua, F.; Berger, A.J.; Cerussi, A.E.; Jakubowski, D.; Tromberg, B.J.
\newblock Broadband Absorption Spectroscopy in Turbid Media by Combined
  Frequency-Domain and Steady-State Methods.
\newblock {\em Appl. Opt.} {\bf 2000}, {\em 39},~6498--6507.
\newblock {{https://doi.org/10.1364/AO.39.006498}}.

\bibitem[O'Sullivan \em{et~al.}(2012)O'Sullivan, Cerussi, Cuccia, and
  Tromberg]{osullivanDiffuseOpticalImaging2012}
O'Sullivan, T.D.; Cerussi, A.E.; Cuccia, D.J.; Tromberg, B.J.
\newblock Diffuse Optical Imaging Using Spatially and Temporally Modulated
  Light.
\newblock {\em J. Biomed. Opt.} {\bf 2012}, {\em 17},~0713111.
\newblock {{https://doi.org/10.1117/1.JBO.17.7.071311}}.

\bibitem[Vasudevan \em{et~al.}(2020)Vasudevan, Forghani, Campbell, Bedford, and
  O'Sullivan]{vasudevanMethodQuantitativeBroadband2020}
Vasudevan, S.; Forghani, F.; Campbell, C.; Bedford, S.; O'Sullivan, T.D.
\newblock Method for {{Quantitative Broadband Diffuse Optical Spectroscopy}} of
  {{Tumor-Like Inclusions}}.
\newblock {\em Appl. Sci.} {\bf 2020}, {\em 10},~1419.
\newblock {{https://doi.org/10.3390/app10041419}}.

\bibitem[Blaney \em{et~al.}(2021)Blaney, Curtsmith, Sassaroli, Fernandez, and
  Fantini]{blaneyBroadbandAbsorptionSpectroscopy2021}
Blaney, G.; Curtsmith, P.; Sassaroli, A.; Fernandez, C.; Fantini, S.
\newblock Broadband Absorption Spectroscopy of Heterogeneous Biological Tissue.
\newblock {\em Appl. Opt.} {\bf 2021}, {\em 60},~7552--7562.
\newblock {{https://doi.org/10.1364/AO.431013}}.

\bibitem[Sassaroli \em{et~al.}(2019)Sassaroli, Blaney, and
  Fantini]{sassaroliDualslopeMethodEnhanced2019}
Sassaroli, A.; Blaney, G.; Fantini, S.
\newblock Dual-Slope Method for Enhanced Depth Sensitivity in Diffuse Optical
  Spectroscopy.
\newblock {\em J. Opt. Soc. Am. A} {\bf 2019}, {\em 36},~1743--1761.
\newblock {{https://doi.org/10.1364/JOSAA.36.001743}}.

\bibitem[Blaney \em{et~al.}(2020)Blaney, Sassaroli, Pham, Fernandez, and
  Fantini]{blaneyPhaseDualslopesFrequencydomain2020}
Blaney, G.; Sassaroli, A.; Pham, T.; Fernandez, C.; Fantini, S.
\newblock Phase Dual-Slopes in Frequency-Domain near-Infrared Spectroscopy for
  Enhanced Sensitivity to Brain Tissue: {{First}} Applications to Human
  Subjects.
\newblock {\em J. Biophotonics} {\bf 2020}, {\em 13},~e201960018.
\newblock {{https://doi.org/10.1002/jbio.201960018}}.

\bibitem[Fantini \em{et~al.}(1999)Fantini, Hueber, Franceschini, Gratton,
  Rosenfeld, Stubblefield, Maulik, and
  Stankovic]{fantiniNoninvasiveOpticalMonitoring1999}
Fantini, S.; Hueber, D.; Franceschini, M.A.; Gratton, E.; Rosenfeld, W.;
  Stubblefield, P.G.; Maulik, D.; Stankovic, M.R.
\newblock Non-Invasive Optical Monitoring of the Newborn Piglet Brain Using
  Continuous-Wave and Frequency-Domain Spectroscopy.
\newblock {\em Phys. Med. Biol.} {\bf 1999}, {\em
  44},~1543--1563.
\newblock {{https://doi.org/10.1088/0031-9155/44/6/308}}.

\bibitem[Martelli \em{et~al.}(1997)Martelli, Contini, Taddeucci, and
  Zaccanti]{martelliPhotonMigrationTurbid1997}
Martelli, F.; Contini, D.; Taddeucci, A.; Zaccanti, G.
\newblock Photon Migration through a Turbid Slab Described by a Model Based on
  Diffusion Approximation {{II Comparison}} with {{Monte Carlo}} Results.
\newblock \textls[-25]{{\em Appl. Opt.} {\bf 1997}, {\em 36},~4600--4612. https://doi.org/10.1364/AO.36.004600.}


\bibitem[Yang \em{et~al.}(2019)Yang, Yang, Wabnitz, Gladytz, Macdonald,
  Macdonald, and Grosenick]{yangSpatiallyenhancedTimedomainNIRS2019}
Yang, L.; Yang, L.; Wabnitz, H.; Gladytz, T.; Macdonald, R.; Macdonald, R.;
  Grosenick, D.
\newblock Spatially-Enhanced Time-Domain {{NIRS}} for Accurate Determination of
  Tissue Optical Properties.
\newblock {\em Opt. Express} {\bf 2019}, {\em 27},~26415--26431.
\newblock {{https://doi.org/89}}.

\bibitem[Patterson \em{et~al.}(1989)Patterson, Chance, and
  Wilson]{pattersonTimeResolvedReflectance1989}
Patterson, M.S.; Chance, B.; Wilson, B.C.
\newblock Time Resolved Reflectance and Transmittance for the Noninvasive
  Measurement of Tissue Optical Properties.
\newblock {\em Appl. Opt.} {\bf 1989}, {\em 28},~2331--2336.
\newblock {{https://doi.org/10.1364/AO.28.002331}}.

\bibitem[Aronson(1995)]{aronsonBoundaryConditionsDiffusion1995}
Aronson, R.
\newblock Boundary Conditions for Diffusion of Light.
\newblock {\em JOSA A} {\bf 1995}, {\em 12},~2532--2539.
\newblock {{https://doi.org/10.1364/JOSAA.12.002532}}.

\bibitem[Popescu \em{et~al.}(2000)Popescu, Mujat, and
  Dogariu]{popescuEvidenceScatteringAnisotropy2000}
Popescu, G.; Mujat, C.; Dogariu, A.
\newblock Evidence of Scattering Anisotropy Effects on Boundary Conditions of
  the Diffusion Equation.
\newblock {\em Phys. Rev. E} {\bf 2000}, {\em 61},~4523--4529.
\newblock {{https://doi.org/10.1103/PhysRevE.61.4523}}.

\bibitem[Ripoll \em{et~al.}(2000)Ripoll, {Nieto-Vesperinas}, Arridge, and
  Dehghani]{ripollBoundaryConditionsLight2000}
Ripoll, J.; {Nieto-Vesperinas}, M.; Arridge, S.R.; Dehghani, H.
\newblock Boundary Conditions for Light Propagation in Diffusive Media with
  Nonscattering Regions.
\newblock {\em JOSA A} {\bf 2000}, {\em 17},~1671--1681.
\newblock {{https://doi.org/10.1364/JOSAA.17.001671}}.

\bibitem[{El-Wakil} \em{et~al.}(2001){El-Wakil}, Degheidy, Machali, and
  {El-Depsy}]{el-wakilRadiativeTransferSpherical2001}
{El-Wakil}, S.A.; Degheidy, A.R.; Machali, H.M.; {El-Depsy}, A.
\newblock Radiative Transfer in a Spherical Medium.
\newblock {\em J. Quant. Spectrosc. Radiat. Transf.}
  {\bf 2001}, {\em 69},~49--59.
\newblock {{https://doi.org/10.1016/S0022-4073(00)00061-3}}.

\bibitem[Optik()]{optikSphereSpectro150H}
Optik, G.
\newblock {{SphereSpectro 150H}}.
\newblock Available online:  
  \url{https://www.gigahertz-optik.com/en-us/product/spherespectro\%20150h/getpdf/} (\hl{25 October 2022}). %MDPI: please add the accessed date.


\end{thebibliography}
\end{document}